\documentclass[preprint]{revtex4}
\usepackage{graphicx}
\usepackage{subfigure}
\usepackage{multirow}
\usepackage{hyperref}
\usepackage{amsmath,bm,amssymb,amsfonts}
\usepackage{color}
\newcommand{\dalm}{\kern1pt\vbox{\hrule height 0.9pt\hbox{\vrule width 0.9pt
			\hskip 2.5pt\vbox{\vskip 5.5pt}\hskip 3pt\vrule width 0.3pt}\hrule height 0.3pt}
	\kern1pt}
\begin{document}

	\title{{\bf
			Scalarized Hot Neutron Stars Containing Hyperons and $\Delta$-Resonances in Different Evolution Regimes}}
	
	\author{{\bf Fahimeh Rahimi $^{1,2}$ }, {\bf Zeinab Rezaei $^{1,2}$} \footnote{Corresponding author. E-mail:
				zrezaei@shirazu.ac.ir}, and {\bf Adamu Issifu $^{3,4}$}}
	\affiliation{ $^{1}$Department of Physics, College of Science, Shiraz University, Shiraz 71454, Iran.\\
		$^{2}$Biruni Observatory, College of Science, Shiraz University, Shiraz 71454, Iran.\\
		$^{3}$Departamento de F\'isica, Instituto Tecnol\'ogico de Aeron\'autica, DCTA, 12228-900, S\~ao Jos\'e dos Campos, SP, Brazil \\
		$^{4}$Laborat\'orio de Computa\c c\~ao Cient\'ifica Avan\c cada e Modelamento (Lab-CCAM) }

	
	\begin{abstract}

Scalar-tensor gravity models are among the prime candidates to explain cosmic acceleration, and compact stars provide unique laboratories for testing such theories. Predictions of scalar-tensor gravity in compact stars can be examined
during the evolution of neutron stars. Spontaneous scalarization in relativistic stars is influenced by different properties of stellar matter in various
evolution regimes. In the present study, we investigate the scalarization of neutron stars in different stages of the evolvement.
For this aim, we apply the isentropic equations of state for the neutron star matter including nucleons, hyperons, and $\Delta$-
resonances in neutrino-trapped, neutrino diffusion, and neutrino-transparent stages as well as cold-catalyzed neutron star. Our equations of state are based on the relativistic model within the mean-field approximation. To emphasize the role of scalar-tensor theories in exploring the properties and structure of compact stars, we calculate the structure of neutron stars
with hyperons and $\Delta$-resonances in different snapshots of the neutron star evolution in the scalar-tensor gravity. Our calculations confirm that the neutron star scalarization is affected by the hyperons as well as the $\Delta$-resonances. Moreover, the properties of scalarized neutron stars depend on the stage of the star evolution.

	\end{abstract}
	
	\maketitle
\section{Introduction} \label{sec:intro}

Compact stars with high enough compactness can experience a tachyonic-like instability known as the spontaneous scalarization.
This non-minimal coupling of the scalar field to neutron stars (NSs) has been constrained applying the data obtained by the timing of binary pulsars,
PSRs J0348+0432, J0737-3039A, J1012+5307, J1738+0333, J1909-3744, J1913+1102, J2222-0137, J0737-3039A, J1913+1102, and J2222-0137 \cite{arXiv:2201.03771}.
Spontaneous scalarization as a nonperturbative strong-field effect significantly depends on the coupling function in the
scalar-tensor theory (STT) \cite{arXiv:1604.04175,arXiv:2006.01153}. Therefore, scalarized NSs are interesting astrophysical laboratories for checking the
modified gravities.
One of the most important questions in this regard is how the spontaneous scalarization takes place in NSs.
The Gauss-Bonnet invariant coupling in STT results in the spontaneous scalarization of NSs \cite{arXiv:1711.02080}.
Scalar-gauge coupling in Einstein gravity can also cause spontaneous scalarization in addition to coupling to curvature invariants \cite{arXiv:2105.14661}.
Linear stability analysis verifies that for the profile with planar symmetric shape, the spontaneous scalarization takes place
more simply compared to the spherically symmetric shape \cite{arXiv:1810.12691}.
Tensor-multi-scalar gravity in three-dimensional maximally symmetric space predicts non-topological scalarized NSs which are
more eligible compared to the general relativity (GR) ones \cite{arXiv:2004.03956}.
Besides, it is interesting to study the factors that affect the spontaneous scalarization in NSs.
Scalar field potentials with massive fields alter the spontaneous scalarization of NS and also
the strong effects from the spontaneous scalarization can result in the observable signatures
from both isolated NS and NS in binaries \cite{arXiv:1601.07475}.
The mass term in STT affects the configuration of NS which is scalarized due to a dynamical transition \cite{arXiv:2407.08124}.
In addition, the spontaneous scalarization can be explored via the influence on the astrophysical objects.
In scalar-tensor model with massive scalar field known as asymmetron, the spontaneous scalarization of asymmetron
can decrease the gravitational constant in NSs \cite{arXiv:1508.01384,arXiv:1707.02809}.
In massive STTs, spontaneous scalarization can lower the stability threshold of proto-neutron stars (PNSs), potentially leading to gravitational collapse into a black hole without necessarily producing shock waves or ejecting the stellar envelope.
It is worth stressing that while the mass radius relation alone may not be sufficient to discriminate STT from GR, since suitable values of the coupling parameter $\beta$ can reproduce similar $M(R)$ curves, other observables can provide clear signatures of scalarization. In particular, the scalar charge, which directly couples to the scalar field, can be constrained through binary pulsar timing observations as it modifies orbital decay via dipolar radiation \cite{Damour:1996ke,Freire:2012mg}. Likewise, tidal deformability \cite{Brown:2022kbw}, measurable in binary NS mergers through gravitational wave signals, is significantly altered in the presence of scalar fields. Additional probes include the moment of inertia mass relation \cite{Staykov:2016mbt}, which can be constrained by precision pulsar measurements, and the thermal evolution of NSs, which may also be influenced by scalar effects. These observables, in combination with $M(R)$, offer promising avenues to distinguish STT from GR in the strong-field regime.

Spin-$1/2$ baryon octet, the nucleons and hyperons, can exist in the NS matter.
Hartree-Fock approximation in $\beta$-equilibrated nucleonic matter with hyperon-nucleon potential predicts that
the hyperons can be present in NSs \cite{arXiv:0811.2939}.
Hyperons may appear in dense NS matter as density rises, and their interactions can lead to a first-order phase transition in some models, though the transition's order depends on the equation of state (EoS) and hyperon interactions \cite{arXiv:astro-ph/0005490}.
The maximum mass of NS with hyperons in the core is in agreement with the constraint
from the pulsar PSR J1614-2230 \cite{arXiv:1308.6121,Sulaksono,Jiang:2012hy}. Besides, the presence of
hyperons in the interior of NSs is consistent with the observations of the pulsars PSR J0030+0451 and PSR J0740+6620
as well as the merger event GW170817 \cite{arXiv:2205.15843}.
Relativistic mean-field models are also in agreement with the
gravitational mass constraint for the pulsar PSR J0348+0432 \cite{arXiv:1403.0341}.
Moreover, relativistic mean-field models suggest that the secondary object in the merger event GW190814 could
be interpreted as a rotating hyperonic star \cite{arXiv:2109.07678}. However, the presence of hyperons in NS interiors,
while theoretically plausible, has not yet been firmly established.

The existence of hyperons in NSs has been also explored through the cooling/heating curves \cite{arXiv:1804.00334,arXiv:2102.0756}
and the luminosity of accreting millisecond pulsar SAX J1808 \cite{arXiv:2102.0756}.
However, the enhancement of star cooling due to the hyperons can be moderated by Joule heating which is a consequence
of crustal electric current dissipation \cite{arXiv:2205.14793}.
Moreover, hyperonic degrees of freedom can increase the influence of temperature
on the thermodynamical observable as well as the NS core composition \cite{arXiv:2206.11266}.
Gravitational waves which are produced from the unstable oscillation modes inside NSs can reveal the existence of hyperons within these stars \cite{arXiv:2011.02204,arXiv:2203.03141,arXiv:2212.09875}.
Besides, hyperonic bulk viscosity alters the $r$-modes in NSs and decreases the star angular velocity \cite{arXiv:astro-ph/0602538,arXiv:1911.08407,arXiv:2208.14436}.
Observational data from massive NSs verify that the star gravitational redshift is in agreement
with stiffer hyperonic EoSs \cite{arXiv:astro-ph/0507312}.
The softening of the EoS due to the hyperons results in a back bending phenomenon
with angular momentum loss and spin-up in NSs \cite{arXiv:astro-ph/0311470}.
The repulsive part of interactions in hyperonic matter due to hyperons \cite{arXiv:1412.8686,arXiv:1602.08106,arXiv:1708.01364,arXiv:1906.11722,
arXiv:1406.4332} as well as the quartic terms involving hidden-strangeness vector meson in non-linear relativistic mean-field model \cite{arXiv:1111.6942} can lead to the stiffness of the NS EoSs resulting into a higher maximum NS mass. Besides, the hyperon-scalar-meson coupling \cite{arXiv:2205.10631} and
the exclusive channel for the hyperon-hyperon interaction mediated by hidden strangeness mesons $(\phi)$ \cite{arXiv:2212.02273} cause the stiffening
of the NS EoS.
The problem on the stiffening of the hyperonic EoS, i.e. hyperon-puzzle, can be solved by appropriate adjustment of the meson-hyperon coupling. The momentum dependence of strangeness potentials \cite{arXiv:2402.08329},
the combined repulsive effects of two- and three-body correlations \cite{arXiv:2001.10563},
the hyperon-nuclear interactions with the properties of SU(3) EFT potentials \cite{arXiv:1612.03758},
the EoS stiffening due to the hadron-quark crossover \cite{arXiv:1508.04861},
the higher-derivative models with power-law terms in
modified f(R) gravity \cite{arXiv:1401.4546}, and
the breaking of SU(6) symmetry to a more general flavor SU(3) symmetry \cite{arXiv:2401.07653}
are the solutions for this problem.
Hyperons in NS alter the speed of sound \cite{arXiv:2204.05221} as well as the star deformability and the gravitational wave signal \cite{arXiv:2108.09113}.

According to the observations of GW170817 with the value $197\leq \Lambda \leq 720$ for the dimensionless tidal deformability,
it is needed to involve the $\Delta$-resonances in the interior of the compact stars \cite{arXiv:1808.02207}.
Hartree and Hartree-Fock approximation confirm the appearance of $\Delta$-resonances in nuclear matter
with densities higher than the saturation density \cite{arXiv:1803.03661}.
$\Delta$-resonances in NSs result in the EoS softening at low and intermediate densities
and EoS stiffening at higher densities \cite{arXiv:1803.03661,arXiv:2202.12083,arXiv:2203.00269}.
In density-dependent relativistic Hartree-Fock theory,
the reduced Fock contributions of $\Delta$-resonances resulting from the coupling of the isoscalar mesons lead to
the softening of the EoS \cite{arXiv:1607.04007}.
Repulsive interactions of $\Delta$-resonances in NS matter alter the composition of matter in the core of NSs \cite{arXiv:2203.00269}.
$\Delta$-resonances in NSs lead to the reduction of star radius and tidal deformability \cite{1907.08583,arXiv:2008.06459}.
The hyperon-free stars and dominated with $\Delta$-resonances with $80\%$ of $\Delta$-resonances at the star center can exist \cite{1907.08583,2206.02935}.
However, GW170817 event is in agreement with two mergering stars containing hyperons and $\Delta$-resonances \cite{1904.02006}.
Considering the $\Delta$-resonances and hyperons in NSs with small
coupling between $\Delta$-resonances and vector mesons decreases the mass and radius of NSs \cite{arXiv:1008.0957,arXiv:1803.03661}.
The fraction of $\Delta$-resonances in NS matter tends to decrease in the presence of hyperons, due to competition for baryon population. However, if $\Delta$-resonances do appear, their coupling to the $\omega$-meson can provide additional repulsion, helping to counteract the softening of the EoS caused by hyperons \cite{2206.02935}. Besides, the appearance of $\Delta$-resonances affects the emergence of hyperons \cite{arXiv:1803.03661}.
In NSs, the vector-isovector self interaction for the mesons and magnetic field increase
the $\Delta$-resonances population and decrease the hyperon density \cite{arXiv:2103.09855}.
Lowest mode frequencies of the radial oscillations in NSs as well as the separation between consecutive modes
increase due to the presence of both hyperons and $\Delta$-resonances in NS matter \cite{arXiv:2303.11006}.

Following the birth of a neutrino-rich PNS, trapped neutrinos in star can play important roles in
the early stages of the evolution. The bulk viscosity peak \cite{arXiv:2406.08978}, the star composition and population of hyperons \cite{arXiv:astroph0209068},
and the star maximum mass \cite{arXiv:nuclth0312050} are affected by the trapped neutrinos. The high-energy neutrinos then diffuse from the supernovae
and lead to the cooling of NS \cite{arXiv:1501.02615,arXiv:1705.06752,arXiv:1705.02122}. In the next stage, the neutrino-transparent
regime forms and all neutrinos exit from the star core (almost 50 seconds after the supernova explosion).
Afterwards, the thermal radiation has significant contribution in the star cooling and eventually, the cold-catalyzed NS forms (nearly 100 years
after the supernova explosion). The evolution of NS should be considered in studying the scalarized NSs with hyperons and $\Delta$-
resonances.

In the present work, we apply the
STT to investigate the scalarization of evolving NSs containing hyperons and $\Delta$-
resonances. The remaining sections are arranged as follows. In Section \ref{s2}, the scalar-tensor gravity which is employed in this work is described
and the Tolman-Oppenheimer-Volkoff equations in this gravity are presented. In Section \ref{s3}, we explain the relativistic model describing the hadronic matter and the NS EoSs in various evolution regimes.
Section \ref{s4} presents our results related to the scalarized NSs. Finally, Section \ref{s5} gives the Summary and Conclusions.

\section{Generalized Tolman-Oppenheimer-Volkoff Equations in Scalar-Tensor Gravity}\label{s2}

In the present work, we apply the formalism presented in Ref. \cite{arXiv:1604.04175} to describe the structure of scalarized compact stars in the STT. We start with the action in Einstein frame in geometric units,
	\begin{equation} \label{action}
		S[g_{\mu \nu},\Phi,\Psi_m] = \frac{1}{16 \pi} \int d^4x \sqrt{-g}(R-2 \nabla_\mu\Phi \nabla^\mu \Phi)+S_m[\Psi_m,a(\Phi)^2 g_{\mu \nu}],
	\end{equation}
with Einstein metric $g_{\mu \nu}$. Here, $g=det(g_{\mu \nu})$, and $R$, $\Phi$, $\Psi_m$ are Ricci scalar, scalar field, and matter field, respectively.
Moreover, $a(\Phi)$ denotes the coupling function and it fixes a particular STT theory. In the Einstein frame, the scalar field couples minimally to the metric. This is while the matter field, $\Psi_m$, couples universally to the conformally rescaled Jordan metric $\tilde{g} _{\mu \nu}=a(\Phi)^2 g_{\mu \nu}$. The above action can be written in the Jordan frame in terms of $\tilde{g} _{\mu \nu}$ leading to a non-minimal coupling of the scalar field and the tensor sector. It is convenient to define $\alpha(\Phi)=\frac{d ln a(\Phi)}{d \Phi}$ and write the following expansion around the scalar field at the present time, $\Phi_0$,
\begin{equation}
\alpha(\Phi)=\alpha_0+\beta_0(\Phi-\Phi_0)+O(\Phi-\Phi_0)^2.	
\end{equation}
The Jordan-Brans-Dicke gravity is obtained considering the first term of the above expansion \cite{Brans1961} with the constrained value $|\alpha_0|\lesssim 3.4\times10^{-3}$. In addition, the next term, $\beta_0$, is the important term in strong-field regime. Considering the values $\beta_0\lesssim -4.35$,
a transition can take place in the relativistic stars leading to the emergence of the scalar field and scalar charge in the stars \cite{arXiv:1604.04175}.
This phenomena is called spontaneous scalarization and can be detected by the observations of compact objects in the binary stars via the enhancement of the energy loss in the emission of gravitational wave due to the extra scalar channel \cite{arXiv:1604.04175}. It should be noted that the scalar-tensor theories with $\beta_0 > 0$ can also result in the spontaneous scalarization \cite{Mendes2015}.

The action in Eq. (\ref{action}) is varied respect to the metric $g_{\mu \nu}$ and the scalar field $\Phi$ to obtain the field equations,
\begin{equation} \label{E1}
	G_{\mu \nu}-2 \nabla_\mu\Phi \nabla_\nu \Phi+g_{\mu \nu}\nabla_\rho\Phi \nabla^\rho \Phi=8\pi a^2 \tilde{T}_{\mu \nu},
\end{equation}
\begin{equation} \label{E2}
	\nabla^\mu \nabla_\mu\Phi=-4\pi a^4 \alpha \tilde{T},
\end{equation}
with the stress-energy-momentum tensor of the matter fields, $\tilde{T}_{\mu \nu}$, and $\tilde{T}=\tilde{g} _{\mu \nu}\tilde{T}^{\mu \nu}$.
In the following, we explore the Einstein frame metric considering the fluid in the Jordan frame. Here, we choose the form $a(\Phi) = e^{\frac{1}{2}\beta (\Phi-\Phi_0) ^2}$ for the coupling function where $\beta$ is the coupling constant and $\Phi_0=0$. The parameter $\beta$ controls the strength of the non-minimal coupling between the scalar field and matter. Negative values of $\beta$ are of particular interest since they can trigger spontaneous scalarization in compact stars \cite{Damour:1993hw,Damour:1996ke}. For this reason, we restrict our study to $\beta=-6,-5$, and $-4.5$, values that are widely adopted in the literature and ensure that scalarization emerges in NSs at astrophysically relevant densities \cite{Doneva:2013qva,Barausse:2012da}. When $\beta=0$, the scalar field decouples from matter and the theory reduces to GR. Very small negative values of $\beta$ suppress scalarization, while very large (negative) values would produce strong deviations from GR already excluded by binary pulsar timing constraints \cite{Antoniadis:2013pzd}.

The stress-energy-momentum tensor of the spherically symmetric perfect fluid describing the NS matter is given by,
\begin{equation}
	\tilde{T}^{\mu \nu}=\tilde{\epsilon}\tilde{u}^{\mu}\tilde{u}^{\nu}+\tilde{p}(\tilde{g}^{\mu \nu}+\tilde{u}^{\mu}\tilde{u}^{\nu}),
\end{equation}
in which $\tilde{ \epsilon}$ and $\tilde{p}$ are the energy density and pressure of the fluid that measured from the observer comoving with the fluid with 4-velocity $\tilde{\textbf{u}}$. Besides, the EoS of the stellar matter, $\tilde{p}(\tilde{ \epsilon})$, should be added to the above equations. The EoSs employed in the present work will be described in the following. For the spherically symmetric static star, we consider the line element,
\begin{equation}
	ds^2 = - N(r)^2 dt^2 + A(r)^2 dr^2 + r^2 (d\theta^2 + \sin^2\theta d\varphi^2),
\end{equation}
in which $N(r)$ and $A(r)$ are metric functions where $A(r) = [1-2 m(r)/r ]^{-1/2}$ and $m(r)$ is the mass profile.
For the static limit, Eqs. (\ref{E1}) and (\ref{E2}) result in the following equations \cite{arXiv:1604.04175},
	\begin{align}
		&\frac{d m}{dr} = 4\pi r^2 a^4 \tilde{\epsilon} + \frac{r}{2} (r-2m) \Big(\frac{d\Phi}{dr}\Big)^2 \label{eq:dm},\\
		&\frac{d \ln N}{dr} = \frac{4\pi r^2 a^4 \tilde{p}}{r - 2m} +\frac{r}{2} \Big(\frac{d\Phi}{dr}\Big)^2 + \frac{m}{r(r-2m)} \label{eq:dn}, \\
		&\frac{d^2\Phi}{dr^2} = \frac{4\pi r a^4}{r-2m} \! \left[ \alpha (\tilde{\epsilon} - 3\tilde{p}) + r (\tilde{\epsilon} - \tilde{p}) \frac{d\Phi}{dr} \right ]\! -\frac{2(r-m)}{r(r-2m)} \frac{d\Phi}{dr} \label{eq:dphi}, \\
		&\frac{d\tilde{p}}{dr} = -(\tilde{\epsilon} + \tilde{p}) \left[  \frac{4\pi r^2 a^4 \tilde{p}}{r-2m} \! + \! \frac{r}{2} \Big(\frac{d\Phi}{dr}\Big)^2 \!\! + \! \frac{m}{r(r-2m)} \! + \! \alpha \frac{d\Phi}{dr} \right], \label{eq:dp}
			\end{align}
known as generalized Tolman-Oppenheimer-Volkoff equations of hydrostatic equilibrium. The boundary conditions which are applied to solve the equations are,
\begin{align}
		&m(0)= 0, \quad \lim_{r\to\infty}N(r) = 1,\quad \Phi(0)=\Phi_c, \quad \lim_{r\to\infty}\Phi(r) = 0, \nonumber \\
		&\dfrac{d\Phi(r)}{dr}(0) = 0, \qquad \tilde{p}(0) = p_c, \qquad \tilde{p}(R_s) = 0,    \label{eq:bc}
	\end{align}
with the star radius $R_s$ and the notation $c$ shows the center of the star. For solving the equations, the solutions in the stellar interior should be matched to the exterior analytical solutions at the surface of the star. For this aim, one can start with a guess $\Phi(0)=\Phi_c$ at the center and do an iteration on $\Phi_c$ in order to satisfying the following condition at the surface of the star,
\begin{equation} \label{eq:constraint on central of scalar field}
		\Phi_s  + \frac{2 \psi_s}{\sqrt{\dot{\nu}_s^2+4\psi_s^2}} \textrm{arctanh} \left[ \frac{\sqrt{\dot{\nu}_s^2 +4\psi_s^2}}{\dot{\nu}_s +2/R_s} \right] = 0,
	\end{equation}
in which the notation s shows the surface of the star and $\psi_s = (d\Phi/dr)_s$, $\dot{\nu}_s = 2(d\ln N/dr)|_s = R_s \psi_s^2 + 2 m_s/[R_s(R_s-2m_s)]$.
This boundary condition is the result of matching the interior and exterior solutions. This boundary condition guarantees the regularity of the scalar field at the center of the star and ensures a smooth matching between the interior solution and the asymptotic exterior solution of the scalar field equations. More details about this boundary condition are available in Refs. \cite{26,arXiv:1604.04175}.
The total mass-energy of an isolated gravitating system at any instant of time measured within a spatial surface enclosing
the system at infinity is called Arnowitt-Deser-Misner (ADM) mass \cite{Baumgarte}. The general exterior solution gives the ADM mass as follows,
\begin{align}
		M_{ADM} &= \frac{R_s^2 \dot{\nu}_s}{2} \left( 1-\frac{2m_s}{R_s} \right)^\frac{1}{2}
		\exp \left[ \frac{-\dot{\nu}_s}{\sqrt{\dot{\nu}_s^2+4\psi_s^2}} \textrm{arctanh} \left( \frac{\sqrt{\dot{\nu}_s^2+4\psi_s^2}}{\dot{\nu}_s +2/R_s} \right) \right].
	\end{align}
In fact, $M_{\rm ADM}$ represents the total gravitational mass of the star as measured at spatial infinity, and it includes contributions from both the matter and scalar field sectors \cite{Arnowitt:1962hi}. Moreover, considering the asymptotic behavior of the scalar field at spatial infinity \cite{26},
\begin{align}
		\Phi=\Phi_0+\omega/r+O(1/r^2),
	\end{align}
introduces the scalar charge, $\omega$. The general exterior solution also determines the scalar charge,
	\begin{align} \label{omM}
				\omega & = - 2 M_{ADM} \psi_s/\dot{\nu}_s.
	\end{align}
The scalar charge, $\omega$, quantifies the asymptotic strength of the scalar field generated by the star and is related to the value of the scalar field at the stellar center \cite{Damour:1993hw}. In this case, $\omega$ provides a measure of the degree of scalarization, which is reflected in the deviations of the mass radius relations compared to the GR case \cite{Pani:2014jra}. In  Eq. (\ref{omM}), the relation between the scalar charge and $M_{\rm ADM}$ is explicitly stated, showing how the scalar field modifies the effective gravitational mass. In the following, we employe the generalized Tolman-Oppenheimer-Volkoff equations to investigate the scalarized NSs in the STT.

\section{Thermodynamical Properties of Hot Neutron Star Matter with Hyperons and $\Delta$-Resonances}\label{s3}

This section will discuss the relevant thermodynamic conditions necessary for studying the evolution of NSs from birth as lepton-rich PNSs to maturity as a cold-catalyzed neutrino-transparent NS. The EoS that forms the basis of the study is calculated from the relativistic mean-field model with density-dependent parameterization which will be discussed in detail in subsequent subsections. In this model, the strong interaction between the baryons is mediated by mesons. The EoS is determined for nucleons (protons ($p$) and neutrons ($n$)) only (N), nucleons plus hyperons ($p,\,n,\,\Lambda,\,\Sigma^{0,\pm},{\rm and},\,\Xi^{0,-}$) (NH), and nucleons plus hyperons plus $\Delta$-resonances ($p,\,n,\,\Lambda,\,\Sigma^{0,\pm},\,\Xi^{0,-},{\rm and} \,\Delta^{0,\pm, ++}$) (NH$\Delta$) similar to the studies carried out in \cite{Issifu:2023qyi,arXiv:2202.12083,Raduta:2021xiz,Issifu:2024fuw}. In this way, one can determine the effect of each particle's degrees of freedom on the NS's evolution.

\subsection{Relativistic Model within the Mean-Field Approximation}
The relativistic mean-field approximation (RMF) is based on the Walecka model \cite{Walecka:1974qa,walecka1986electroweak} or the so-called quantum hadrodynamics which incorporates the two fundamentally known interactions inside the nucleus: attractive contribution for long-distance separation of particles and repulsive for short distances separation of particles. The Yukawa-type potentials representing the scalar and the vector mesons that mediate the interaction introduce these effects at different distance scales (for a review, see \cite{Menezes:2021jmw} and references therein). Generally, the Lagrangian density for the RMF is given by

\begin{equation}\label{eq}
     \mathcal{L}_{\rm RMF}= \mathcal{L}_{H}+ \mathcal{L}_{\Delta}+ \mathcal{L}_{\rm m}+ \mathcal{L}_{\rm L},
\end{equation}
where $\mathcal{L}_{H}$ represents the baryon octet, $\mathcal{L}_{\Delta}$ the $\Delta$-resonances, $\mathcal{L}_{\rm m}$ the mesons and the leptons are represented by $\mathcal{L}_{\rm L}$. The baryon octet consists of spin-1/2 particles that a Dirac-type Lagrangian can describe
\begin{align}\label{a}
 \mathcal{L}_{H}= {}& \sum_{b\in H}  \bar \psi_b \Big[  i \gamma^\mu\partial_\mu - \gamma^0  \big(g_{\omega b} \omega_0  +  g_{\phi b} \phi_0+ g_{\rho b} I_{3b} \rho_{03}  \big) - \Big( m_b- g_{\sigma b} \sigma_0 \Big)  \Big] \psi_b,
\end{align}
the summation is over all the spin-1/2 baryons present, $b\in (p,\,n,\,\Lambda,\,\Sigma^{0,\pm},\,\Xi^{0,-})$.
With a scalar meson $\sigma$ and vector mesons $\omega$ and $\phi$, while the latter vector meson carries a hidden strangeness, they are both isoscalars in nature and $\Vec{\rho}$ is a vector-isovector meson and $g_{ib}$ ($i = \sigma,\, \omega,\, \phi,\, {\rm and},\, \rho$) are the meson-baryon couplings. The subscript `0' of the meson fields in the Lagrangian indicates their mean-field versions. On the other hand, the leptons are represented by free Dirac-type Lagrangian
\begin{equation}\label{l1}
    \mathcal{L}_{\rm L} = \sum_L\Bar{\psi}_L\left(i\gamma^\mu\partial_\mu-m_L\right)\psi_L,
\end{equation}
where the summation runs over electrons $e$, muons $\mu$, and their corresponding neutrinos ($\nu_e\,{\rm and}\, \nu_\mu$). Additionally, the $\Delta$-resonances are spin-3/2 particles, hence they have an additional vector-valued spinor projection compared to the baryon octet and the leptons, therefore they are described by the Rarita-Schwinger-type Lagrangian
\begin{align}\label{a1}
        \mathcal{L}_{\Delta}={}& \sum_{d\in \Delta}\Bar{\psi}_{d}\Big[i \gamma^\mu \partial_\mu- \gamma^0\left(g_{\omega d}\omega_0 + g_{\rho d} I_{3d} \rho_{03} \right) -\left(m_d-g_{\sigma d}\sigma_0 \right)\Big]\psi_{d}.
\end{align}
Nonetheless, studies show that the equation of motion of the spin-3/2 particles can be written in the same way as the spin-1/2 particles in the RMF approximation regime \cite{dePaoli:2012eq}. Finally, the mesonic part that mediates the strong interactions in the RMF Lagrangian can be expressed as
\begin{align}\label{a11}
 \mathcal{L}_{\rm m}= - \frac{1}{2} m_\sigma^2 \sigma_0^2  +\frac{1}{2} m_\omega^2 \omega_0^2  +\frac{1}{2} m_\phi^2 \phi_0^2 +\frac{1}{2} m_\rho^2 \rho_{03}^2,
\end{align}
where $m_i$ are the meson masses whose values are presented in Tab.~\ref{T}. Here, we have neglected the derivative terms of the meson fields because they vanish under the mean-field approximation.

\subsection{Density-Dependent Parametrization}

\begin{table}
\begin{center}
\begin{tabular}{ |c| c| c| c| c| c| c| }
\hline
 meson($i$) & $m_i(\text{MeV})$ & $a_i$ & $b_i$ & $c_i$ & $d_i$ & $g_{i N} (n_0)$\\
 \hline
 $\sigma$ & 550.1238 & 1.3881 & 1.0943 & 1.7057 & 0.4421 & 10.5396 \\
 $\omega$ & 783 & 1.3892 & 0.9240 & 1.4620 & 0.4775 & 13.0189  \\
 $\rho$ & 763 & 0.5647 & --- & --- & --- & 7.3672 \\
 \hline
\end{tabular}
\caption {DDME2 parameters \cite{PhysRevC.71.024312}.}
\label{T}
\end{center}
\end{table}

\begin{table}
\begin{center}
\begin{tabular}{ | c | c | c | c | c | }
\hline
 b,d & $\chi_{\omega b,d}$ & $\chi_{\sigma b,d}$ & $\chi_{\rho b,d}$ & $\chi_{\phi b}$  \\
 \hline
 $\Lambda$ & 0.714 &  {0.646} & 0 & -0.808  \\
$\Sigma^0$ & 1 &  {0.663} & 0 & -0.404  \\
  $\Sigma^{-}$, $\Sigma^{+}$ & 1 &  {0.663} & {1} & -0.404  \\
$\Xi^-$, $\Xi^0$  & 0.571 & {0.453} & 0 &  {- 1.01} \\
  $\Delta^-$, $\Delta^0$, $\Delta^+$, $\Delta^{++}$   & 1.285  & {1.331} & 1 & 0  \\
  \hline
\end{tabular}
\caption {The ratio of the baryon coupling to the corresponding nucleon coupling for hyperons and $\Delta$s \cite{Lopes:2022vjx}.}
\label{T1}
\end{center}
\end{table}
In the density-dependent parameterizations scheme, the baryon-meson couplings are determined as a function of baryon density $n_B$. This eliminates the effect of temperature and fluctuations in the effective interactions of the particles in the theory. In this work, we adopt the improved density-dependent nucleon-meson couplings (DDME2) \cite{PhysRevC.71.024312} given by the expression
\begin{equation}
    g_{i b} (n_B) = g_{ib} (n_0)a_i  \frac{1+b_i (\eta + d_i)^2}{1 +c_i (\eta + d_i)^2},
\end{equation}
with $i=\sigma, \omega, \phi$ and
\begin{equation}
    g_{\rho b} (n_B) = g_{\rho b} (n_0) \exp\left[ - a_\rho \big( \eta -1 \big) \right],
\end{equation}
where $\eta =n_B/n_0$.
The parameters of the model are determined by fitting to the experimental data of bulk nuclear matter properties around its saturation density ($n_0 = 0.152\,\rm fm^{-3}$), binding energy ($B/A= -16.40\,\rm MeV$), the compressibility modulus ($K_0 = 251.09\,\rm MeV$), symmetry energy ($J = 32.30\, \rm MeV$) and slope ($L_0 = 51.30\,\rm MeV$) \cite{Dutra:2014qga,Reed:2021nqk,Lattimer:2023rpe}. The nucleon-meson coupling constants $g_{iN}$ and the model parameters determined at $n_0$ are presented in Tab.~\ref{T}. There are several ways to extend the nucleon-meson couplings to include other particles such as the hyperons and the $\Delta$-resonances in the literature. Some of the most employed approaches are universal coupling \cite{PhysRevLett.67.2414}, fixed and varying potential depths \cite{PhysRevC.85.065802,Weissenborn:2011kb,Lopes:2013cpa,PhysRevC.88.015802}, and symmetry and non-symmetry arguments \cite{Lopes:2022vjx,Cavagnoli:2011ft}. We adopt the RMF model calibrated with the DDME2 \cite{PhysRevC.71.024312} parameterization, as it accurately reproduces nuclear saturation properties and finite nuclei \cite{Typel:2005ba}, while also yielding NS maximum masses and radii in agreement with current astrophysical observations \cite{Fortin:2016hny}. Its density-dependent couplings provide a robust framework for extrapolation to the supranuclear densities encountered in NSs. Furthermore, this choice allows us to isolate the effects of thermodynamic conditions and the influence of scalar-tensor theories, given its demonstrated consistency with both experimental data and astrophysical measurements.

In this work, we adopt the couplings determined in \cite{Lopes:2022vjx} using SU(3) and SU(6) symmetry arguments. Aside from the authors calculating the meson-baryon couplings for spin-1/2 octet, they extended their calculations to determine the meson-baryon coupling for the spin-3/2 decuplet in a model-independent manner. This minimizes the arbitrariness in determining the $\Delta$-meson couplings. The only particle whose potential depth is well known in the literature is $\Lambda$, $U_\Lambda=-28\,\rm MeV$, hence, the authors used it as the foundation to determine $U_\Xi=-4\,\rm MeV$, $U_\Sigma=30\,\rm MeV$ and $U_\Delta\approx-98\,\rm MeV$ that formed the bases for calculating the couplings. However, the determination is done relative to the  $g_{iN}$, such that $\chi_{ib,d} = g_{ib,d}/g_{iN}$. In Tab.~\ref{T1} we represent the values of the couplings used for this work, this corresponds to $\alpha_V=0.5$ in \cite{Lopes:2022vjx}. The parameter $\alpha_V$ governs the relative strength of the vector-meson couplings to hyperons and $\Delta$-resonances \cite{Lopes:2022vjx}. Our chosen value, $\alpha_V = 0.5$, represents an optimized choice that satisfies constraints from hypernuclear physics while providing sufficient repulsion to produce a NS EoS consistent with observed maximum masses in the presence of the exotic baryons. This choice of parameterization has been used in \cite{Issifu:2023qyi} to study the evolution of NSs with exotic baryons, hybrid NSs in \cite{Issifu:2023ovi} and PNSs with quark core in \cite{Issifu:2024fuw}. Therefore, this parameterization has been tested and it has proven to produce stars that satisfy the $2\,\rm M_\odot$ threshold determined by NS observational data \cite{NANOGrav:2019jur,Fonseca:2021wxt}.

\subsection{Neutron Star Matter Equation of State}
The thermodynamic quantities required to calculate the EoS are determined from the Lagrangian densities presented in (\ref{eq}) to (\ref{a11}). The baryon densities  are given as
\begin{equation}\label{a2}
n_b = \gamma_b \int \frac{d^3 k}{(2\pi)^3}  \left[f_{b\,+}(k) - f_{b\,-}(k)  \right],
\end{equation}
where $\gamma_b = 2$ is the baryon degeneracy and $f_{\rm b \pm}(k)$ represents the Fermi-Dirac distribution function, expressed explicitly as
\begin{equation}
    f_{b \pm}(k) = \frac{1}{1+\exp[(E_b \mp \mu^\ast_b)/T]}, \nonumber
\end{equation}
with the single particle energy $E_b = \sqrt{k^2+m_b^*}$, where $m_b^*$ is the effective mass of the baryon and $T$ is temperature. This expression also holds for the $\Delta$-resonances when we substitute $b\leftrightarrow d$, $E_d = \sqrt{k^2+m_d^*}$, and $\gamma_d = 4$ due to the additional spin projection. The effective chemical potential $\mu^*$ that appears in $f_{\rm b \pm}(k)$ can be expressed explicitly in terms of $b$ and $d$ as
\begin{align}
    \mu_{b,d}^\ast &= \mu_{b,d}- g_{\omega {b,d}} \omega_0 - g_{\rho {b,d}} I_{3{b,d}} \rho_{03} - g_{\phi {b}} \phi_0 - \Sigma^r,
\end{align}
where $\mu_{b,d}$ are the real chemical potentials and $\Sigma^r$ is the so-called rearrangement term \cite{Typel:1999yq,Fuchs:1995as} which is determined through thermodynamic consistency analysis
\begin{align}
    \Sigma^r ={}& \sum_b \Bigg[ \frac{\partial g_{\omega b}}{\partial n_b} \omega_0 n_b + \frac{\partial g_{\rho b}}{\partial n_b} \rho_{03} I_{3b}  n_b+ \frac{\partial g_{\phi b}}{\partial n_b} \phi_0 n_b - \frac{\partial g_{\sigma b}}{\partial n_b} \sigma_0 n_b^s + b\leftrightarrow d\Bigg].
\end{align}
The associated effective masses are
\begin{equation}\label{a3}
    m_{b,d}^\ast =m_{b,d}- g_{\sigma b,d} \sigma_0, 
\end{equation}
while the scalar density of the baryons are
\begin{equation}
    n_{b}^s =\gamma_b \int \frac{d^3 k}{(2\pi)^3} \frac{m^\ast_b}{E_b} \left[f_{b\,+}(k) + f_{b\,-}(k)  \right].
\end{equation}
It also holds for the $\Delta$-resonances when we interchange $b\leftrightarrow d$, with the appropriate degeneracy factor. Similarly, it holds for leptons with degeneracy of $\gamma_L =2$, for $e$ and $\mu$, with a constant chemical potential $\mu_e$. On the contrary, when we consider neutrinos in the stellar matter, then $\gamma_L =1$ for the neutrinos. The equations of motion of the Lagrangian densities are
\begin{equation}\label{eom1}
  m^2_\sigma \sigma_0 = \sum_b g_{\sigma b}n^s_b + \sum_d g_{\sigma d}n_d^s,
\end{equation}

\begin{equation}\label{eom2}
m^2_\omega\omega_0 = \sum_b g_{\omega b}n_b + \sum_d g_{\omega d}n_d,
\end{equation}

\begin{equation}\label{eom3}
m^2_\phi \phi_0 = \sum_b g_{\phi b}n_b,
\end{equation}

\begin{equation}\label{eom4}
m^2_\rho\rho_{03} = \sum_b g_{\rho b}n_bI_{3b} + \sum_d g_{\rho d}n_dI_{3d},
\end{equation}
we can numerically calculate the meson fields which leads to determining, $\mu^*$, $m^*$, and temperature fluctuations in the matter. Consequently, the energy density and the pressure of the system can be determined through the energy-momentum tensor trace yielding
\begin{flalign}\label{1a}
    \varepsilon_B&=  \varepsilon_b + \varepsilon_m + \varepsilon_d +\varepsilon_L,\\
    P_B&=  P_b + P_m +P_d + P_L + P_r\label{1b},
\end{flalign}
the subscripts are representative of the individual particles in the system, while $P_r$, is the pressure correction arising from the thermodynamic consistency of the model. The energy density and pressure contributions of the fermions to the total energy density $\varepsilon_B$ and pressure $P_B$ are
\begin{equation}\label{eq:ener_b}
    \varepsilon_b= \gamma_b \int \frac{d^3 k}{( 2\pi)^3} E_b \left [f_{b+}(k) +f_{b-}(k) \right],
\end{equation}
and
\begin{equation}\label{eq:press_b}
    P_b= \gamma_b \int \frac{d^3 k}{( 2\pi)^3} \frac{k^2}{E_b} \left [ f_{b+}(k) +f_{b-}(k) \right],
\end{equation}
respectively. The same expressions hold for the $\Delta$-resonances and the leptons replacing $b$ with $d$ and $L$ with the corresponding degeneracies. On the other hand, the contribution from the meson fields is given by
\begin{equation}\label{eq:ener_m}
    \varepsilon_m=  \frac{m_\sigma^2}{2} \sigma_0^2+\frac{m_\omega^2}{2} \omega_0^2 +\frac{m_\phi^2}{2} \phi_0^2  + \frac{m_\rho^2}{2} \rho_{03}^2,
\end{equation}
and
\begin{equation}\label{eq:press_m}
    P_m= -  \frac{m_\sigma^2}{2} \sigma_0^2 +\frac{m_\omega^2}{2} \omega_0^2 +\frac{m_\phi^2}{2} \phi_0^2  + \frac{m_\rho^2}{2} \rho_{03}^2.
\end{equation}
The pressure correction is related to $\Sigma^r$ as
\begin{equation}
    P_r = n_B\Sigma^r,
\end{equation}
which can be determined through
\begin{equation}
    P_B = n_B^2\dfrac{\partial }{\partial n_B}\left( \dfrac{\varepsilon_B}{n_B}\right).
\end{equation}
Following the thermodynamic quantities discussed above, we can determine the entropy per baryon of the system through the Helmholtz free energy expression $\mathcal{F}_B= \varepsilon_B - Ts_B$, where $s_B=S/n_B$ is the entropy per baryon and $S$ the entropy density of the system. Therefore, we can write
\begin{equation}
     s_B = \frac{\varepsilon_B +P_B-
     \sum_b \mu_b n_b -
     \sum_d \mu_d n_d
}{T},
\end{equation}
which served as the basis for calculating the EoSs of the stellar matter. In this relation, we can calculate the temperature fluctuations of the stellar matter using the EoS for fixed values of $s_B$.

For completeness, the detailed derivation of the $T=0$ case can be found in \cite{Menezes:2021jmw,Typel:1999yq}. However, in the limit $T=0$,
\begin{equation}
    f_{b_\pm}(k) = \frac{1}{1+\exp\!\left[{(E_b \mp \mu_b^*})/{T}\right]}
    \;\xrightarrow[T=0]{}\;
    \Theta(\mu_b^* - E_b) ,
\end{equation}
where $\Theta$ denotes the step function. At this limit, the effective chemical potential becomes equal to the Fermi energy, $\mu_b^*(T=0)=E_F$. This substitution correctly reproduces the expressions for the number density, pressure, and energy density in the $T=0$ case.

\subsection{Stellar Matter for Hot NSs}
The EoS composed of N, NH, and NH$\Delta$ matter was calculated separately by solving the equations of motion (\ref{eom1})-(\ref{eom4}) of the particles together with $n_{b,d}$ and $n^s_{b,d}$ to determine the meson fields and the corresponding chemical potentials for each composition of the stellar matter. We fixed the $s_B$ and the lepton fraction $Y_{L,e}=(n_e+n_{\nu_e})/n_B$ to calculate the hot EoSs. The predominant leptons in newly born NS are known to be electrons and their corresponding neutrinos according to supernova physics \cite{Prakash:1996xs,PhysRevC.100.015803}. Nonetheless, the muons only become relevant after neutrinos have diffused from the stellar matter, hence, we set $Y_{L,\mu}=(n_\mu+n_{\nu_\mu})/n_B \approx 0$, at the initial stages of the star's evolution while the $\tau$-leptons are considered too heavy to be present. Different combinations of $s_B$ and $Y_{L,e}$ are chosen to calculate the EoSs that accurately describe the evolution of the NS from birth as a neutrino-rich object to the formation of a cold-catalyzed neutrino-poor object based on the relevant simulation data in the literature \cite{PhysRevC.100.015803,Shao:2011nu,Nakazato:2021yjv,Raduta:2020fdn}.

We consider four different stages of stellar evolution: two stages for neutrino-rich matter and two stages for neutrino-poor stellar matter. In the first stage, we consider a newly born NS with $s_B = 1$ ($s_B$ is in the unit of Boltzmann constant $k_B$ which is set to unity along the text) and $Y_{L,e}=0.4$ (neutrino-trapped stage). The second stage is when the star is about $0.5 - 1s$ old, here the entropy per baryon is estimated to be between $1\leq s_B \leq 3$, while the $Y_{L,e}$ decreases from the first stage due to neutrino diffusion, hence, we choose $s_B = 2$ and $Y_{L,e}=0.2$ (neutrino diffusion stage). At the third stage, all the neutrinos have completely escaped from the stellar core when the star is about $\sim 15s$ old, at this stage the star gets maximally heated and will start to cool down as the $s_B$ decreases, we choose $s_B = 2$ and $Y_{\nu_e} = 0$ for this stage (neutrino-transparent stage). The last stage considered in the series of snapshots analyses is when the star catalyzes ($s_B = 0$, this is expected to occur when the star is $\sim 50s$ old) and cools down over several years through thermal radiations to form a cold `mature' NS, at this stage the thermodynamic properties are $T=0$ and $Y_{\nu_e} = 0$ (cold-catalyzed stage). These thermodynamic properties of the NS matter from birth as neutrino-rich PNSs to `maturity' as cold-catalyzed neutrino-transparent NS are discussed in detail in \cite{Janka:2006fh,Prakash:1996xs}.

Furthermore, to establish equilibrium between all the particle species in the stellar matter due to weak processes, the following chemical potential relations are used;
\begin{align}
    &\mu_\Lambda = \mu_{\Sigma^0} = \mu_{\Xi^0} = \mu_{\Delta^0} = \mu_{n}=\mu_B,\\
    &\mu_{\Sigma^-} = \mu_{\Xi^-} = \mu_{\Delta^-} = \mu_{B}-\mu_Q,\\
    &\mu_{\Sigma^+} = \mu_{\Delta^+} = \mu_p=  \mu_{B}+\mu_Q, \\
    &\mu_{\Delta^{++}}=\mu_{B}+2\mu_Q,
\end{align}
where $\mu_B$ is the baryon chemical potential and $\mu_Q = \mu_p-\mu_n$ is the charge chemical potential. The subscripts of $\mu$ represent the individual particle species under consideration. However, in neutrino-transparent matter $\mu_Q = -\mu_e$, with $\mu_e=\mu_\mu$, when we consider the presence of neutrinos trapped in the stellar matter, $\mu_Q$ is modified to include the presence of neutrinos, therefore
\begin{equation}
    \mu_Q = \mu_{\nu_e} - \mu_e.
\end{equation}
It is important to mention that the electron neutrinos are sufficient for studying trapped neutrinos in PNSs in supernova physics. The stellar matter is also charge neutral, hence
\begin{equation}
n_p+n_{\Sigma^+}+2n_{\Delta^{++}}+n_{\Delta^+} -(n_{\Sigma^- }+ n_{\Xi^-} + n_{\Xi^-} + n_{\Delta^-})=n_Q,
\end{equation}
where $n_Q = n_e + n_\mu$.

\begin{figure}[h]
	\subfigure{}{\includegraphics[scale=0.72]{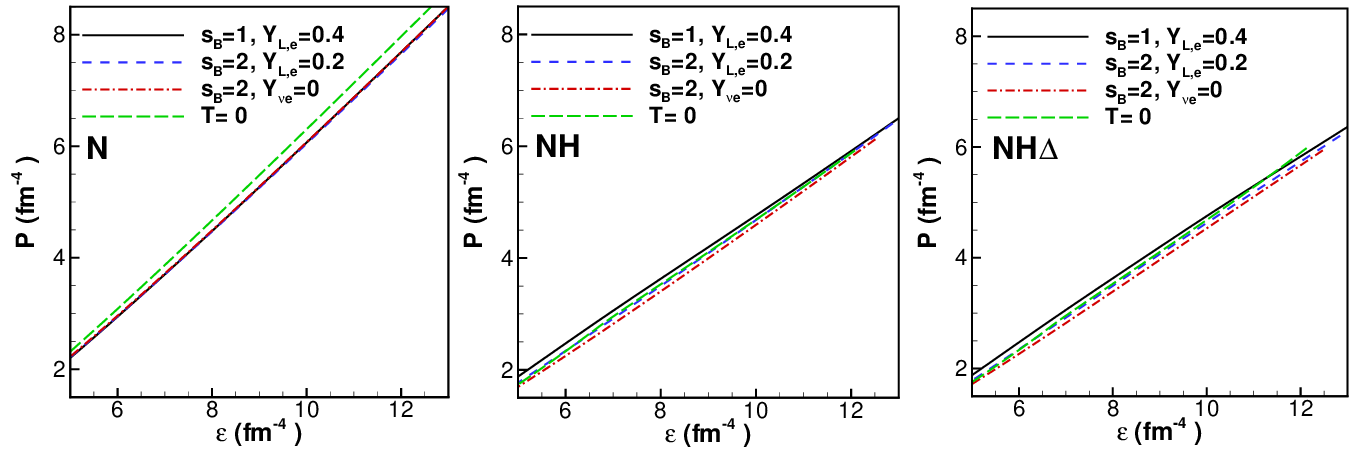}
		}	
	\caption{Neutron star matter equation of state in four stages of the star's evolution, Neutrino-Trapped Regime ($s_B = 1, Y_L = 0.4$), Neutrino Diffusion Regime ($s_B = 2, Y_L = 0.2$), Neutrino-Transparent Regime ($s_B = 2, Y_{\nu_e} = 0$), and Cold-Catalyzed Regime (T=0) in three cases with N, NH, and NH$\Delta$ matter.}
	\label{EOS-regime2}
\end{figure}

\begin{figure}[h]
	\subfigure{}{\includegraphics[scale=1]{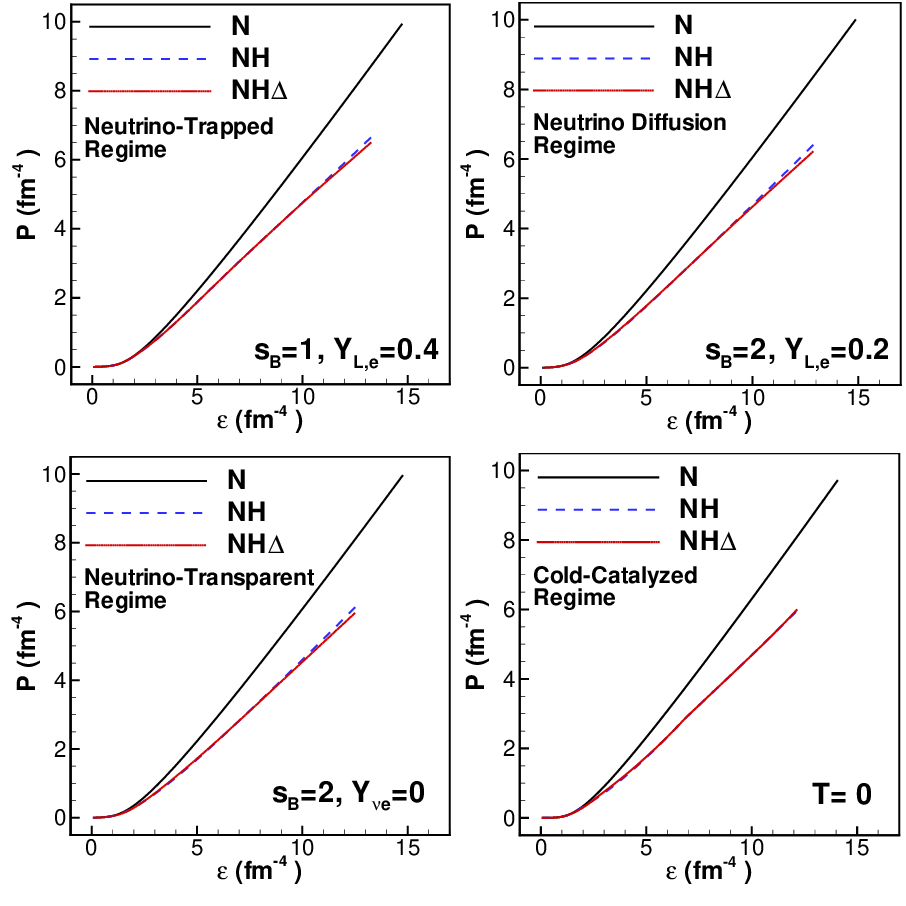}
		}	
	\caption{Comparison of neutron star matter equation of state containing N, NH, and NH$\Delta$ matter for different regimes of the star evolution.}
	\label{p}
\end{figure}

Fig.~\ref{EOS-regime2} shows the NS EoSs considering different phases of the evolution for three cases of N, NH, and NH$\Delta$ matter. In the case of N matter,
the EoSs of neutrino-trapped, neutrino diffusion, and neutrino-transparent regimes are almost similar. This is while the EoS of cold-catalyzed regime is stiffer than the other ones. For the cases of NH and NH$\Delta$ matter, the effects of the star evolution on the EoS stiffening are different from N matter.
For NH and NH$\Delta$ matter, the EoS of neutrino-trapped phase is the stiffest one. The EoS of the star matter in the neutrino diffusion regime is softer than the EoS of the previous regime, i.e. neutrino-trapped one. Moreover, in neutrino-transparent regime, the EoS becomes softer and actually the EoS of neutrino-transparent regime is the softest EoS. In the next phase, i.e. the cold-catalyzed one, the EoS becomes stiffer compared to the neutrino diffusion and neutrino-transparent regimes, while it is still softer than the EoS of neutrino-trapped regime. Thus, across the panels, the stiffest EoS for the fixed-entropy stars appears in the first stage, when the star is relatively cold and lepton-rich, and in the final stage, when it becomes cold and catalyzed. This highlights the role of entropy and lepton fractions in shaping the EoS. The second and third stages show a softer EoS, which naturally follows as neutrinos escape the stellar core and the temperature peaks. The softening results from the increased thermal pressure, which expands the stellar radius and thereby reduces the central pressure.

In Fig.~\ref{p}, we also compare the EoSs of N, NH, and NH$\Delta$ matter for the four stages of the NS evolution considered in this study. We use different colors on the curves to differentiate the particle composition of each EoS. We observe that the EoS composed of nucleons (solid black line) constitutes the stiffer EoS, as the degrees of freedom of the matter increase by including hyperons (blue dashed line) and $\Delta$-resonances (solid red line) the EoS gradually softens. Consequently, the maximum expected stellar mass will also reduce with increasing degrees of freedom of the stellar matter. Since stiffer EoSs lead to higher maximum masses and vice versa. From the recent measurement of the millisecond pulsar PSR J0740+6620 through the NICER observatory, with a measured maximum mass $2.072^{+0.067}_{-0.066}\rm M_\odot$ and radius $12.39^{+1.30}_{-0.98}\rm km$, simultaneously determined at 68\% confidence level \cite{Fonseca:2021wxt,Riley:2021pdl,LIGOScientific:2018cki} and the gravitational wave detection in event GW170817 \cite{LIGOScientific:2017vwq,LIGOScientific:2018cki}, gives a clear mass radius threshold for describing NSs. These observational data pose a constraint on the EoSs describing NSs. Since the free parameters of the RMF model are determined through fitting to experimental nuclear matter properties, the only way to address these constraints within the model framework is by adjusting the baryon-meson couplings of the non-nucleonic constituents of the stellar matter \cite{Weissenborn:2011ut,Lopes:2022vjx,2206.02935}. The choice of parametrization in this work is carefully selected to ensure that irrespective of the composition of the NS, the $2\rm M_\odot$, maximum mass threshold is satisfied.

 \begin{figure}[ht]
  \centering
  \includegraphics[scale=0.9]{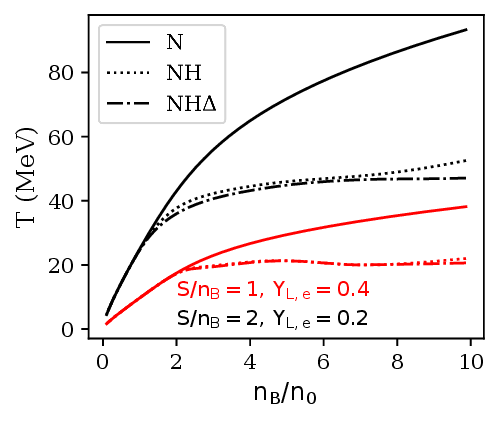}
  \includegraphics[scale=0.9]{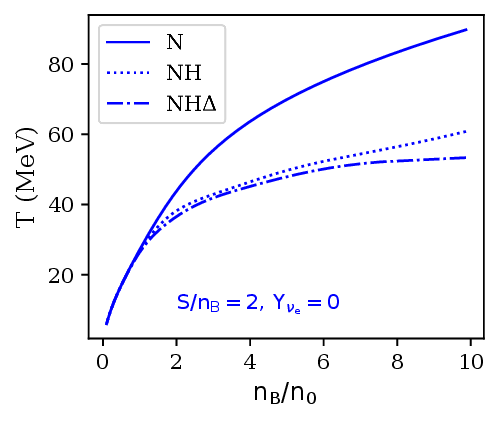}
\caption{The temperature distributions in the stellar matter are shown as a function of the ratio of the baryon density ($n_B$) to the nuclear saturation density ($n_0$). The left panel shows the neutrino-trapped and neutrino diffusion stages and the right panel shows the neutrino-transparent stage.}
 \label{ss1}
\end{figure}

In Fig.~\ref{ss1}, we present the temperature distribution in the stellar matter for the three evolutionary stages with fixed entropy. In general, the temperature increases with rising $s_B$ and decreasing $Y_{L,e}$ as the star evolves. The first stage corresponds to the lowest temperature profile: shortly after core bounce, the star is relatively cold, but shock waves combined with neutrino diffusion (second stage) raise the $s_B$ of the stellar matter, resulting in a temperature increase \cite{1986ApJ...307..178B,Pons:1998mm}. By the third stage, neutrinos have fully diffused out of the stellar core, and the temperature reaches its maximum before cooling down through thermal radiation. The introduction of new degrees of freedom further modifies the thermal behavior \cite{Issifu:2023qyi}. Since hyperons and $\Delta$-resonances provide additional channels to distribute the entropy, they lower the thermal energy per particle, which in turn reduces the net temperature of the system. Consequently, across the panels, nucleonic matter exhibits the highest temperature distribution, while the inclusion of hyperons reduces it, and the additional presence of $\Delta$-resonances lowers it even further. These thermal effects have direct consequences for the stellar structure. An increase in temperature softens the EoS, leading to an expansion of the star and a reduction in compactness. Thus, both the thermal evolution and the appearance of exotic degrees of freedom play key roles in determining the stability and structure of PNSs \cite{Glendenning2000}.

\section{Scalarization of Evolving Neutron Stars with Hyperons and $\Delta$-Resonances}\label{s4}

\subsection{Neutron Star Mass}

\begin{figure}[h]
	\subfigure{}{\includegraphics[scale=0.55]{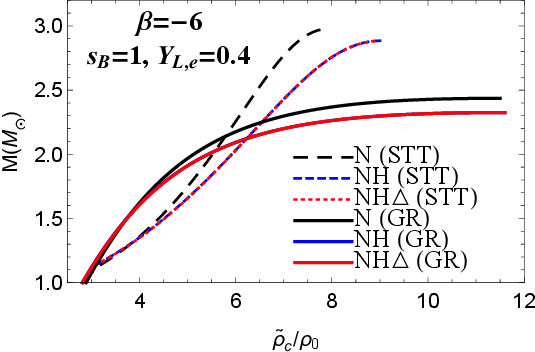}
		}	
\subfigure{}{\includegraphics[scale=0.55]{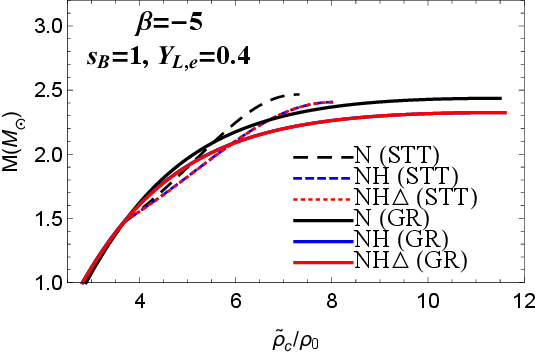}
		}	
\subfigure{}{\includegraphics[scale=0.55]{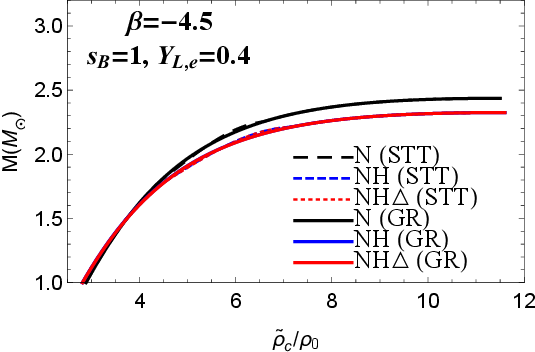}
		}	
	\subfigure{}{\includegraphics[scale=0.55]{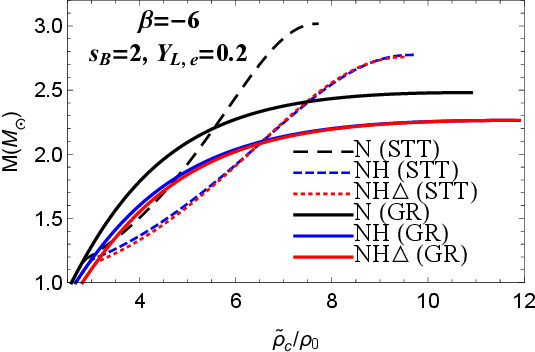}
		}	
\subfigure{}{\includegraphics[scale=0.55]{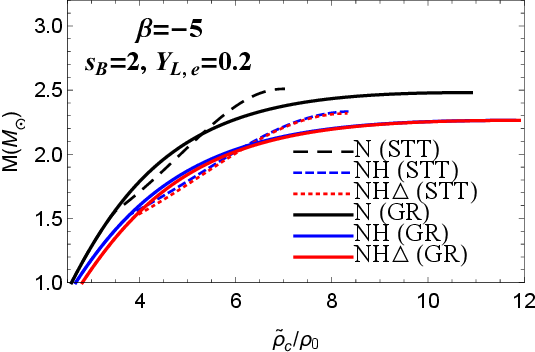}
		}	
\subfigure{}{\includegraphics[scale=0.55]{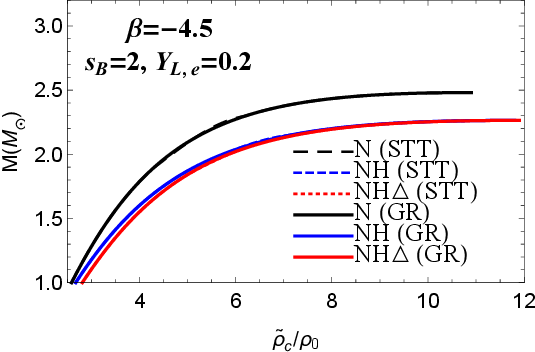}
		}
\subfigure{}{\includegraphics[scale=0.55]{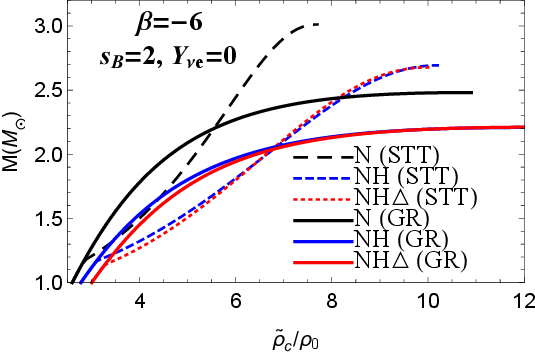}
		}	
\subfigure{}{\includegraphics[scale=0.55]{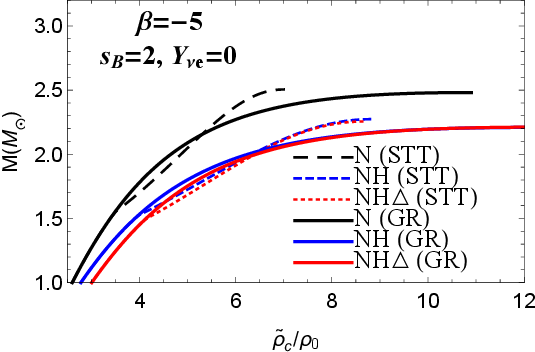}
		}	
\subfigure{}{\includegraphics[scale=0.55]{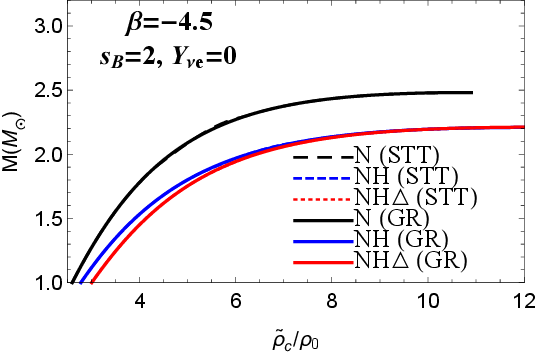}
		}	
	\subfigure{}{\includegraphics[scale=0.55]{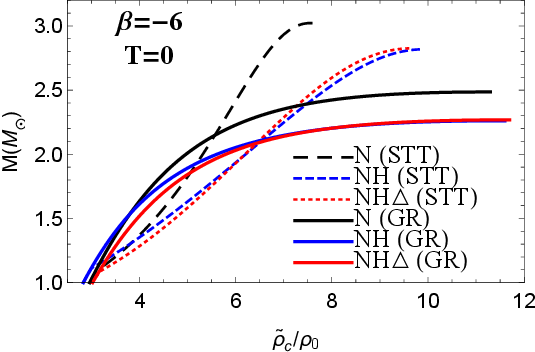}
		}	
\subfigure{}{\includegraphics[scale=0.55]{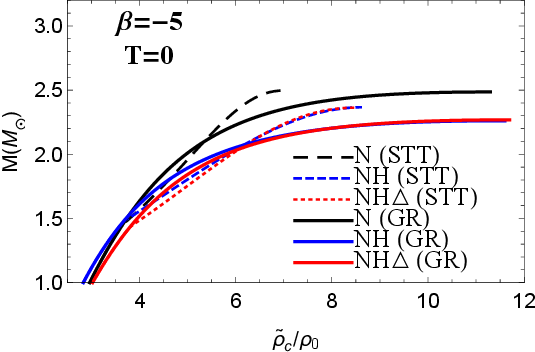}
		}	
\subfigure{}{\includegraphics[scale=0.55]{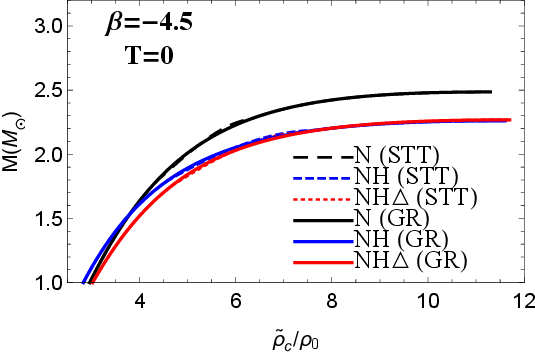}
		}		
\caption{Neutron star mass versus the central density, $\tilde{\rho}_{c}$, considering different contents, i.e. N, NH, and NH$\Delta$ matter, in the scalar-tensor theory (STT) and general relativity (GR) with three values of the coupling constant $\beta$ in four stages of the star's evolution, Neutrino-Trapped Regime ($s_B = 1, Y_L = 0.4$), Neutrino Diffusion Regime ($s_B = 2, Y_L = 0.2$), Neutrino-Transparent Regime ($s_B = 2, Y_{\nu_e} = 0$), and Cold-Catalyzed Regime (T=0). We have considered the value $\rho_0=1.66\times10^{14}g/cm^3$ to present the dimensionless density.}
	\label{mro1}
\end{figure}

Fig.~\ref{mro1} gives the NS mass versus the central density of stars in the cases of N, NH, and NH$\Delta$ matter at different stages of the
star's evolution. Our results for the star mass in the STT can be different from the GR one.
This difference appears in the scalarized stars. The presence of hyperons and $\Delta$-resonances
alters the deviation of scalarized star mass from the GR one. With hyperons and $\Delta$-resonances, scalarization
shifts to higher densities and the stars can be scalarized even with larger densities.
This is due to the EoS softening by the hyperons and $\Delta$-resonances.
The density range in which the scalarization takes place is wider in the cases of hyperons and $\Delta$-resonances.
It is evident that the scalarization reduces (enhances) the mass for low (high) density stars.
The density at which the maximum mass of scalarized stars appears is higher when the hyperons and $\Delta$-resonances
are considered. Hyperons and $\Delta$-resonances reduce the star mass for both scalarized and non-scalarized stars, as a result of the softening of NH and NH$\Delta$ EoS.
In the first stage of the star evolution, i.e. neutrino-trapped regime with $s_B = 1$ and $Y_L = 0.4$, the behavior of
NH star is almost similar to NH$\Delta$ one. However, in the next phases, NH stars are different from NH$\Delta$ ones. In the neutrino diffusion regime with $s_B = 2$ and $Y_L = 0.2$ as well as the neutrino-transparent
regime with $s_B = 2$ and $ Y_{\nu_e} = 0$, the existence of $\Delta$-resonances
decreases the mass of scalarized stars especially in stars with low central density.
This is while for cold-catalyzed NSs with T=0, the presence of $\Delta$-resonances increases the maximum mass of scalarized stars. Moreover, for the low mass stars in the neutrino diffusion, neutrino-transparent, and cold-catalyzed regimes,
the hyperons and $\Delta$-resonances affect the star mass unlike the neutrino-trapped regime.
With the higher values of the coupling constant, the scalarization is suppressed and the results of the STT approach the GR ones.

\begin{figure}[h]
\subfigure{}{\includegraphics[scale=0.55]{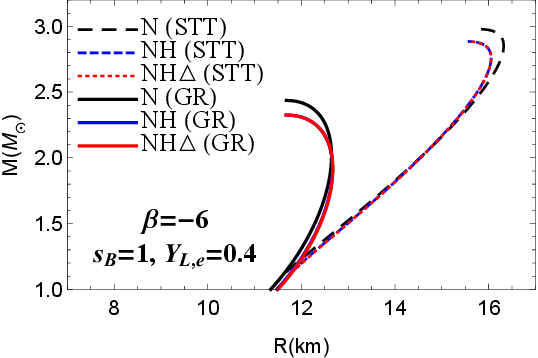}
		}	
\subfigure{}{\includegraphics[scale=0.55]{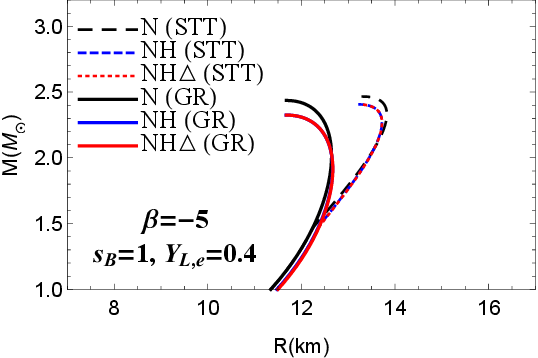}
		}	
\subfigure{}{\includegraphics[scale=0.55]{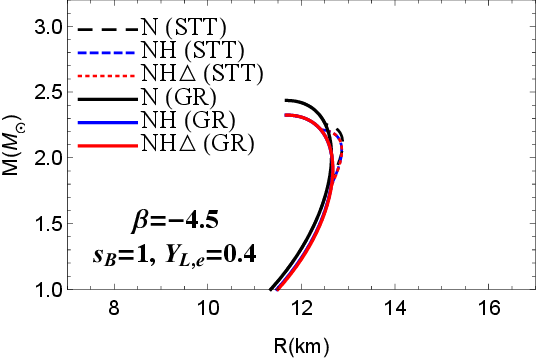}
		}
	\subfigure{}{\includegraphics[scale=0.55]{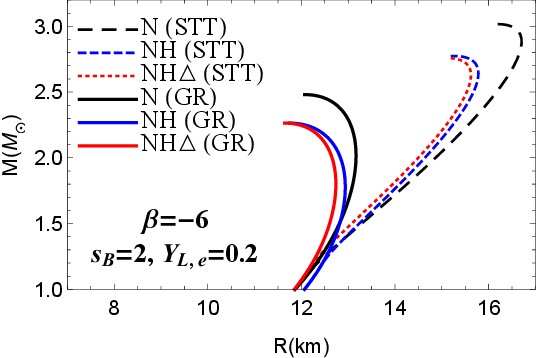}
		}	
\subfigure{}{\includegraphics[scale=0.55]{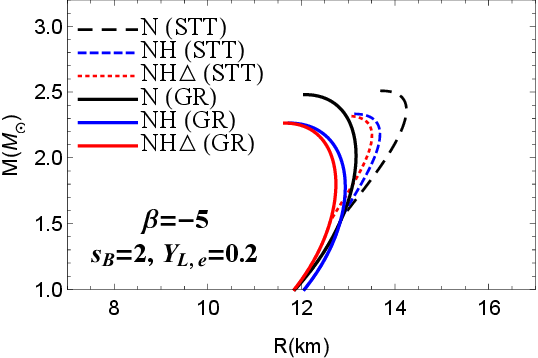}
		}	
\subfigure{}{\includegraphics[scale=0.55]{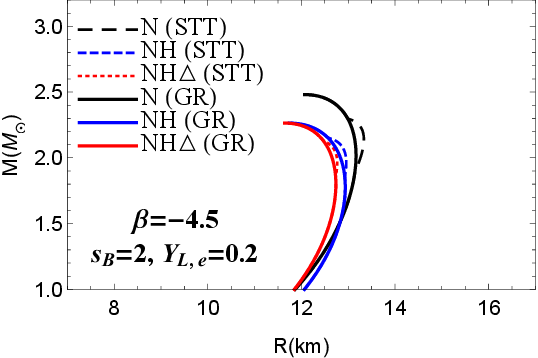}
		}	
\subfigure{}{\includegraphics[scale=0.55]{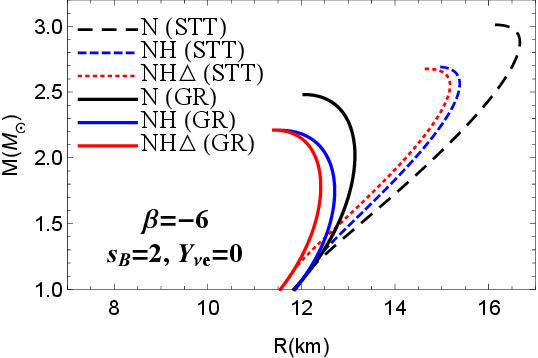}
		}	
\subfigure{}{\includegraphics[scale=0.55]{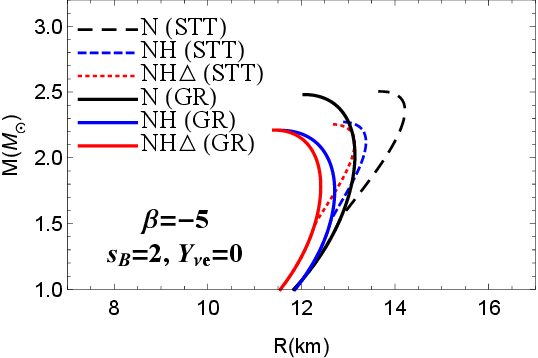}
		}	
\subfigure{}{\includegraphics[scale=0.55]{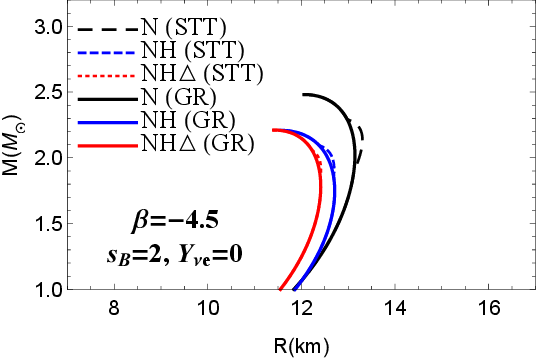}
		}	
	\subfigure{}{\includegraphics[scale=0.55]{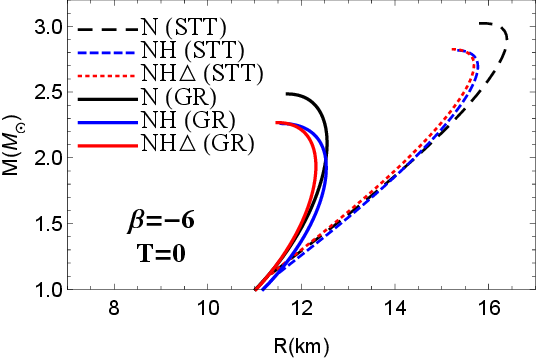}
		}	
\subfigure{}{\includegraphics[scale=0.55]{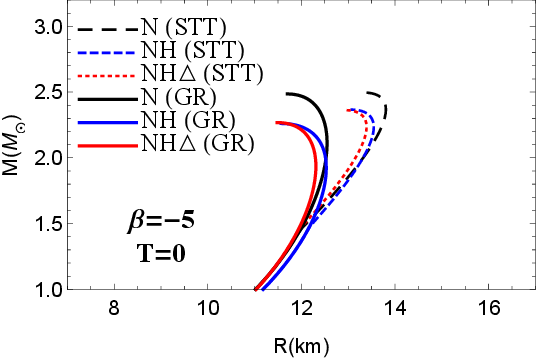}
		}	
\subfigure{}{\includegraphics[scale=0.55]{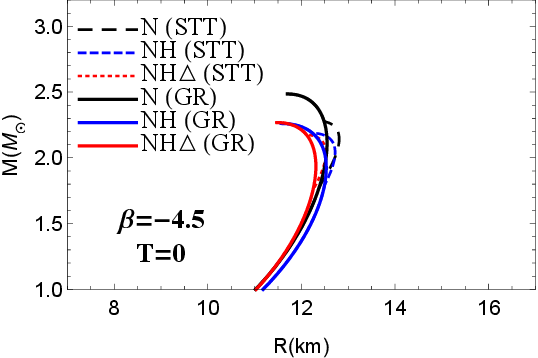}
		}	
	\caption{Same as Fig.~\ref{mro1} but for the mass versus the star radius.}
	\label{M-R}
\end{figure}

In Fig.~\ref{M-R}, we have presented the mass radius relation for the NSs with N, NH, and NH$\Delta$ matter in different
regimes of the evolution. The scalarization of the stars increases both the size and the mass of the stars.
The lower value of the coupling constant amplifies these enhancements. In the scalarized stars, the radius can get the values around 16km (with $\beta=-6$).
For the scalarized stars, the mass radius relation behaves differently from the one in GR. In the scalarized stars
with NH and NH$\Delta$ matter, considering a special star mass, the radius of the star is lower than
the one with N matter. In STT, the star mass is higher than the one in GR and can reach the values
about $3M_\odot$ (with $\beta=-6$). In the neutrino-trapped regime with $\beta=-6$, the mass radius relations for scalarized stars with
N, NH, and NH$\Delta$ matter and the mass lower than the value about $2.5M_\odot$ are nearly the same. However, the mass radius relations
for the stars with NH and NH$\Delta$ matter differ from the one with N matter when the star mass is higher than $2.5M_\odot$. We can also
explore the effects of NH and NH$\Delta$ matter on the mass radius relation of scalarized stars in the next regimes of the star evolution.
Fig.~\ref{M-R} confirms that in the neutrino diffusion, neutrino-transparent, and cold-catalyzed stages, for all scalarized stars even the ones with
low masses, the mass radius relations are different in three cases of N, NH, and NH$\Delta$ matter. Although this difference is more evident for
massive stars. Our results show that the range of the scalarization and the deviation of STT from GR in the mass radius relation depend on the
evolution regime and this range is more extended in cold-catalyzed stars. Moreover, with the lower values of the coupling constant,
the range of the scalarization increases.

\subsection{Neutron Star Scalar field}

\begin{figure}[h]
\subfigure{}{\includegraphics[scale=0.55]{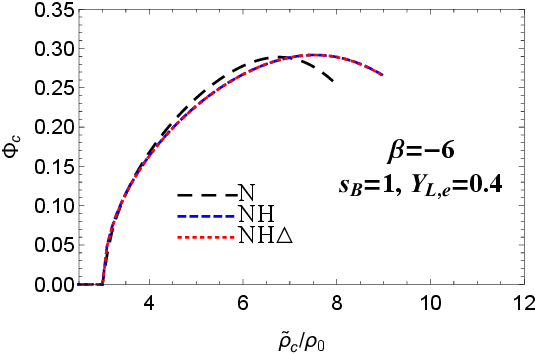}
		}	
\subfigure{}{\includegraphics[scale=0.55]{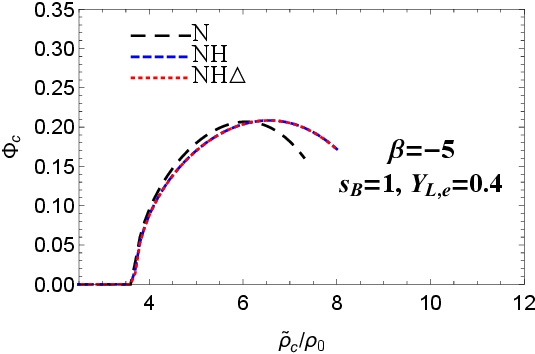}
		}	
\subfigure{}{\includegraphics[scale=0.55]{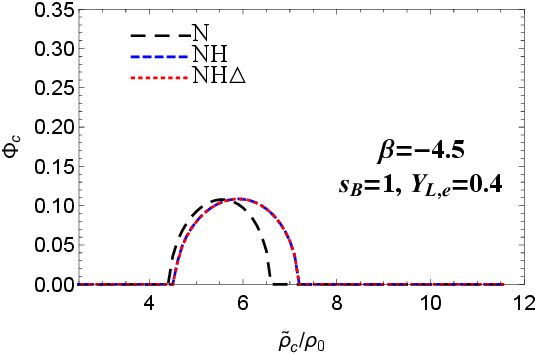}
		}	
	\subfigure{}{\includegraphics[scale=0.55]{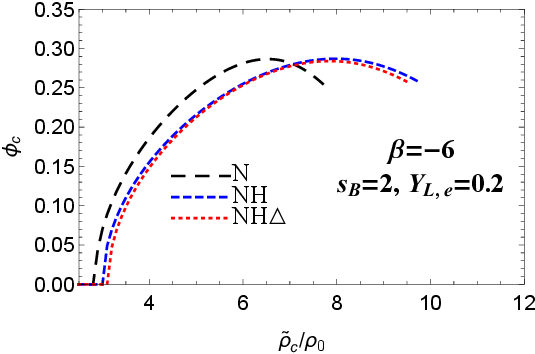}
		}	
\subfigure{}{\includegraphics[scale=0.55]{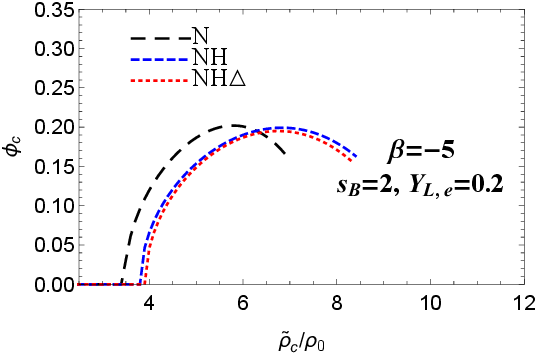}
		}	
\subfigure{}{\includegraphics[scale=0.55]{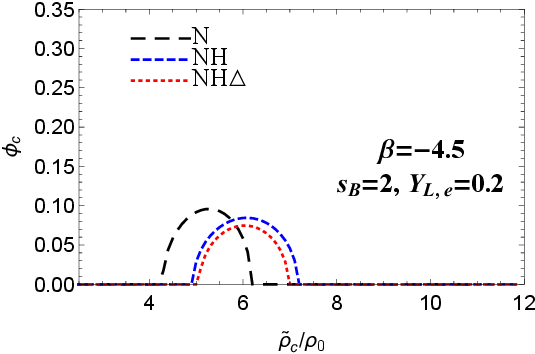}
		}	
\subfigure{}{\includegraphics[scale=0.55]{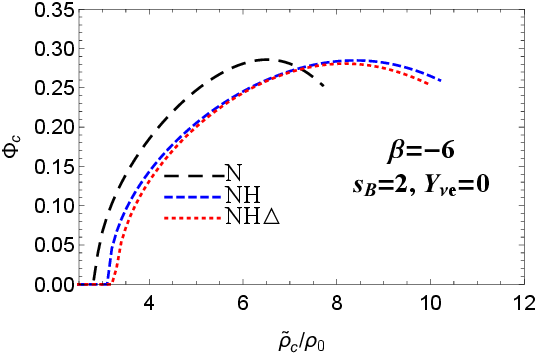}
		}	
\subfigure{}{\includegraphics[scale=0.55]{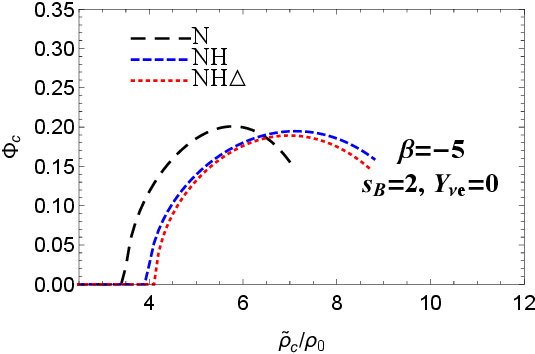}
		}	
\subfigure{}{\includegraphics[scale=0.55]{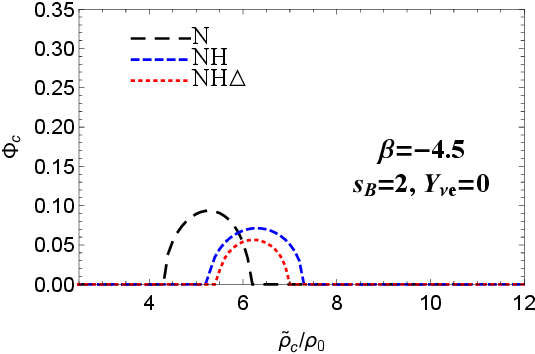}
		}	
	\subfigure{}{\includegraphics[scale=0.55]{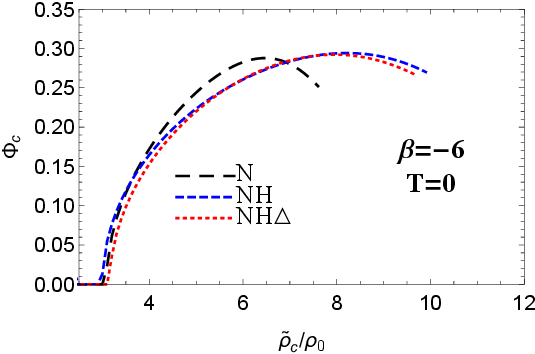}
		}	
\subfigure{}{\includegraphics[scale=0.55]{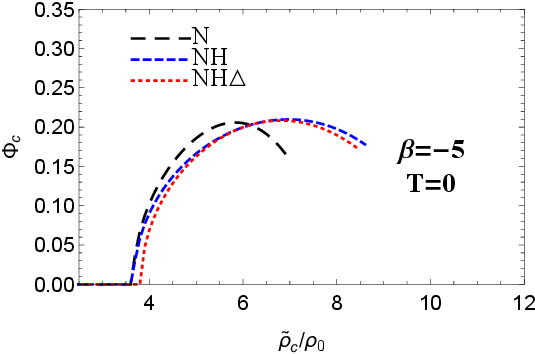}
		}	
\subfigure{}{\includegraphics[scale=0.55]{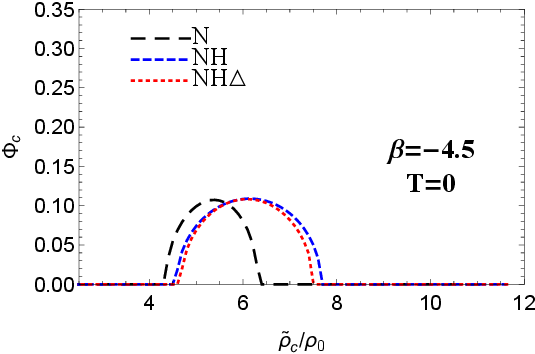}
		}	
	\caption{The values of the scalar field at the center of the star, $\Phi_c$, versus the central density, $\tilde{\rho}_{c}$, for the stars containing N, NH, and NH$\Delta$ matter, with different values of the coupling constant $\beta$ considering various evolution regimes, Neutrino-Trapped Regime ($s_B = 1, Y_L = 0.4$), Neutrino Diffusion Regime ($s_B = 2, Y_L = 0.2$), Neutrino-Transparent Regime ($s_B = 2, Y_{\nu_e} = 0$), and Cold-Catalyzed Regime (T=0).}
	\label{phi1}
\end{figure}

The central scalar field versus the central density of stars with N, NH, and NH$\Delta$ matter
considering the star evolution is presented in Fig.~\ref{phi1}. Scalar field at the center of the stars
with low central densities which are non-scalarized is zero. For the stars with the central density
higher than a special value (first critical density, $\rho_{crit1}$), the central scalar field is nonzero and the stars are scalarized.
This situation continues for higher central densities and the central scalar field grows with the central density.
At a special value of the central density, $\Phi_c$ approaches to a maximum value. This means that the stars with this central density
experience the maximum scalarization. For the stars with higher densities at the center, the central scalar field reduces. Considering the value $\beta=-4.5$ for the coupling constant, $\Phi_c$ decreases to value $\Phi_c=0$ at another special value of the central density (second critical density, $\rho_{crit2}$). For the stars with the central densities higher than the second critical density, $\Phi_c$ is zero and the scalarization disappears. Tabs.~\ref{T5} and ~\ref{T6} present the values of $\rho_{crit1}$ and $\rho_{crit2}$ respectively for the stars in different evolution regimes containing N, NH, and NH$\Delta$ matter. It is clear that in the cases of $\beta=-6$ and $\beta=-5$, the stars remain scalarized up to high densities and there is no second critical density. In the neutrino-trapped regime, the first critical density does not depend on the content of the stars and $\rho_{crit1}$ is almost equal for the
cases of N, NH, and NH$\Delta$ stars. In this regime, however, the second critical density for the
cases of NH and NH$\Delta$ stars is higher than the one in N stars. This is due to the influences of the hyperons and $\Delta$-resonances
on the softening of the EoS. Besides, for the neutrino diffusion and neutrino-transparent stages,
the EoS softening by the hyperons and $\Delta$-resonances grows the first critical density. Tab.~\ref{T5} also verifies that for the cold-catalyzed stage, the softening of the NH$\Delta$ EoS can enhance the value of the first critical density. This is while in neutrino diffusion, neutrino-transparent, and cold-catalyzed regimes, the hyperons enhance the second critical density and the $\Delta$-resonances reduce it. This decrement of $\rho_{crit2}$ by the $\Delta$-resonances can be a result of the EoS stiffening at high densities due to these particles. The effects of hyperons and $\Delta$-resonances are more significant in the neutrino diffusion and neutrino-transparent regimes. This can be attributed to the reduced neutrino pressure support during these stages of stellar evolution, which makes the star more susceptible to gravitational effects. For NH and NH$\Delta$ stars in the neutrino diffusion, neutrino-transparent, and cold-catalyzed regimes, the stars with higher central densities can be scalarized. This means that the range of the scalarization in these three regimes is more extended compared to the neutrino-trapped regime. This is a result of the EoS softening of NH and NH$\Delta$ matter in the neutrino diffusion, neutrino-transparent, and cold-catalyzed regimes compared to the neutrino-trapped EoS (see Fig.~\ref{EOS-regime2}). In all stages of the star evolution, the higher values of $\beta$ assist the differing $\rho_{crit1}$ in
the cases of N, NH, and NH$\Delta$ matter. In addition, the first critical density increases by increasing the coupling constant. Our results verify that the presence of hyperons and $\Delta$-resonances extends the range of nonzero scalar field and the scalarization of the stars.
Moreover, the stars experience the higher values of the scalar field when the coupling constant is lower.

\begin{table}
\begin{center}
\begin{tabular}{ |c | c| c| c| c| c| c|c| c| c|}
\hline   & \multicolumn{1}{c}{} &\multicolumn{1}{c}{$\beta=-6$} &\multicolumn{1}{c|}{}&\multicolumn{1}{c}{} &\multicolumn{1}{c}{$\beta=-5$} &\multicolumn{1}{c|}{}&\multicolumn{1}{c}{} &\multicolumn{1}{c}{$\beta=-4.5$} &\multicolumn{1}{c|}{}\\
\hline
Regime &N & NH & NH$\Delta$   &N &NH  &NH$\Delta$   & N & NH & NH$\Delta$  \\
 \hline
 Neutrino-Trapped &3.1  & 3.1 & 3.1 & 3.7 &3.7  &3.7  & 4.5  &4.6  & 4.6  \\
Neutrino Diffusion &2.9  &3.1  &3.2  & 3.5 &3.9  &4.0  & 4.3  &5.0  &  5.1 \\
  Neutrino-Transparent &2.9  & 3.2 & 3.3 & 3.5 & 4.0 &4.2  & 4.4  & 5.3 &5.5  \\
Cold-Catalyzed &3.1  &3.0  & 3.2 &  3.7 & 3.7 &3.9 & 4.4 & 4.6 & 4.7  \\
  \hline
\end{tabular}
\caption {First critical density of the scalarization, $\rho_{crit1}$, in different regimes of the evolution for the stars with N, NH, and NH$\Delta$ matter at different values of the coupling constant.}
\label{T5}
\end{center}
\end{table}

\begin{table}
\begin{center}
\begin{tabular}{| c|c| c| c|}
\hline Coupling Constant  &\multicolumn{1}{c}{} & \multicolumn{1}{c}{$\beta=-4.5$} &\multicolumn{1}{c|}{}\\
\hline
Regime & N & NH & NH$\Delta$  \\
 \hline
 Neutrino-Trapped & 6.5  & 7.1 & 7.1  \\
Neutrino Diffusion & 6.1  & 7.1 & 6.9  \\
  Neutrino-Transparent &6.1   & 7.2 & 6.9 \\
Cold-Catalyzed &  6.3 &7.6  & 7.5  \\
  \hline
\end{tabular}
\caption {Second critical density, $\rho_{crit2}$, in different regimes of the evolution for the stars with N, NH, and NH$\Delta$ matter with the value of $\beta=-4.5$ for the coupling constant.}
\label{T6}
\end{center}
\end{table}

\subsection{Neutron Star Scalar Charge}

\begin{figure}[h]
	\subfigure{}{\includegraphics[scale=0.55]{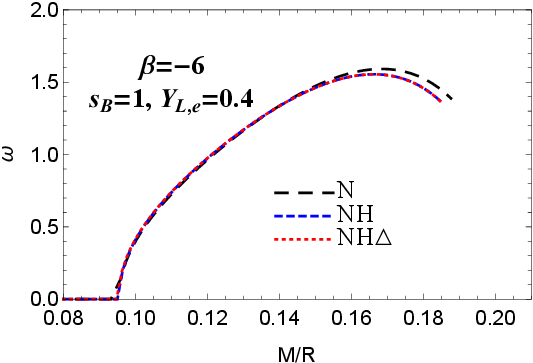}
		}		
\subfigure{}{\includegraphics[scale=0.55]{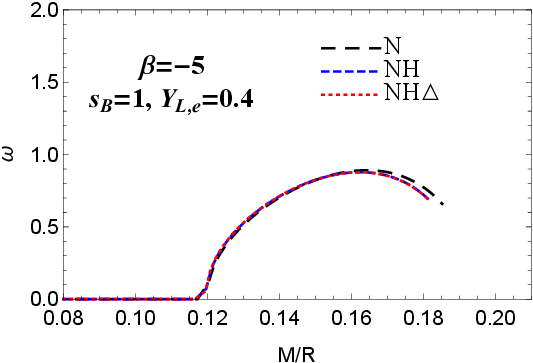}
		}	
\subfigure{}{\includegraphics[scale=0.55]{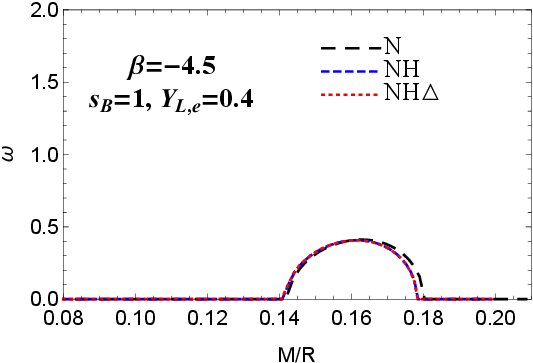}
		}	
\subfigure{}{\includegraphics[scale=0.55]{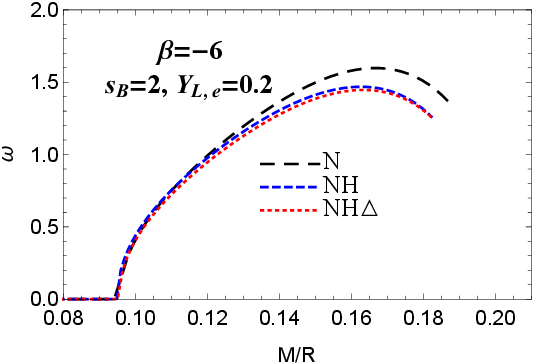}
		}		
\subfigure{}{\includegraphics[scale=0.55]{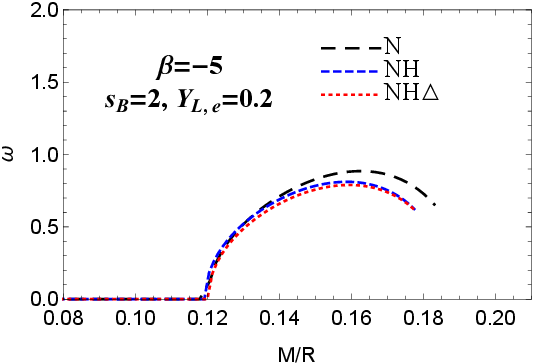}
		}	
\subfigure{}{\includegraphics[scale=0.55]{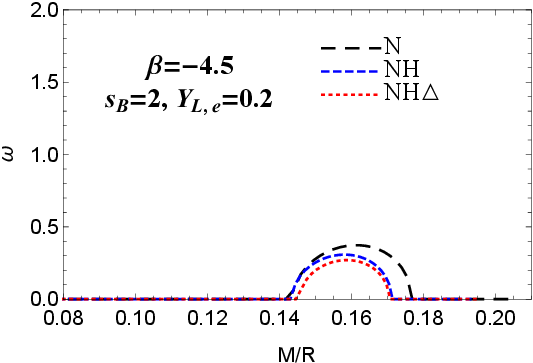}
		}	
	\subfigure{}{\includegraphics[scale=0.55]{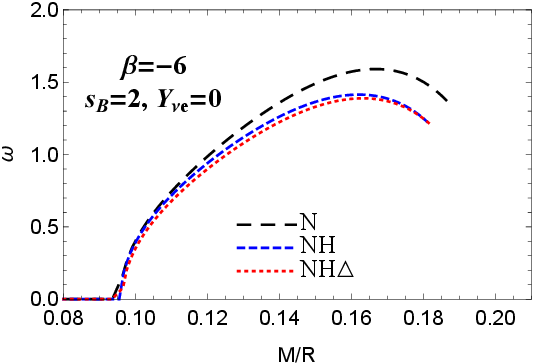}
		}		
\subfigure{}{\includegraphics[scale=0.55]{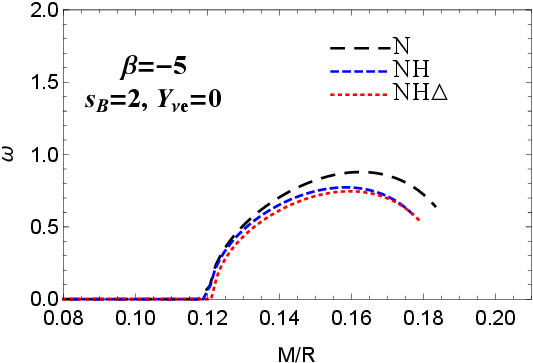}
		}	
\subfigure{}{\includegraphics[scale=0.55]{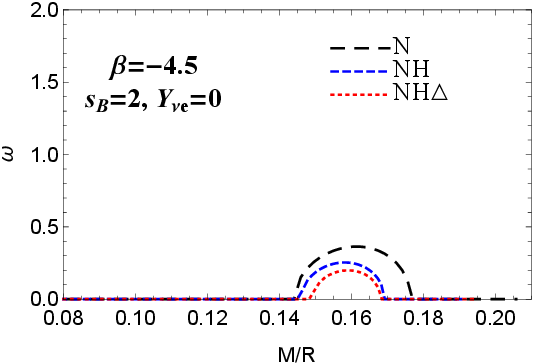}
		}
\subfigure{}{\includegraphics[scale=0.55]{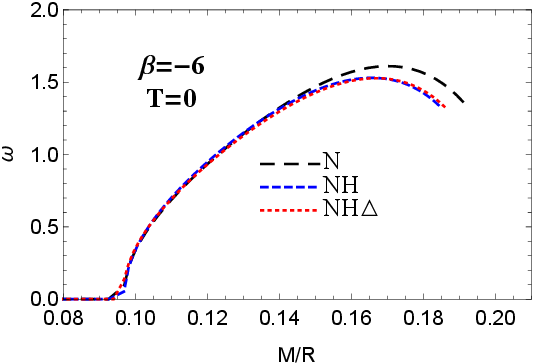}
		}		
\subfigure{}{\includegraphics[scale=0.55]{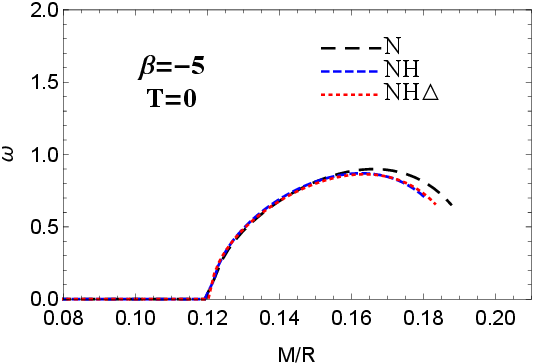}
		}	
\subfigure{}{\includegraphics[scale=0.55]{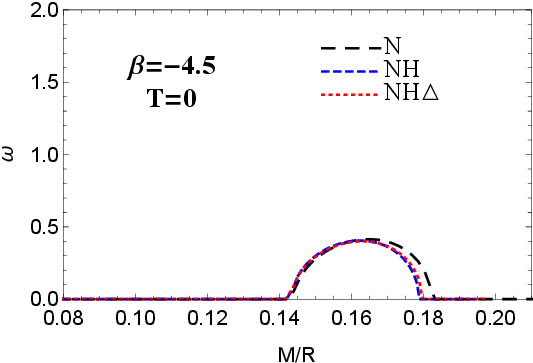}
		}	
	\caption{Scalar charge, $\omega$, as a function of the star compactness, $M/R$, in neutron stars with N, NH, and NH$\Delta$ matter and three values of the coupling constant $\beta$ considering the star's evolution, Neutrino-Trapped Regime ($s_B = 1, Y_L = 0.4$), Neutrino Diffusion Regime ($s_B = 2, Y_L = 0.2$), Neutrino-Transparent Regime ($s_B = 2, Y_{\nu_e} = 0$), and Cold-Catalyzed Regime (T=0).}
	\label{charge1}
\end{figure}

Fig.~\ref{charge1} presents the scalar charge, $\omega$, of NSs in the cases of N, NH, and NH$\Delta$ matter considering
different evolution regimes. The star scalar charge is an indication of the spontaneous scalarization in the star.
It is obvious that the presence of hyperons and $\Delta$-resonances
reduces the star scalar charge, as a result of the softening of the NH and NH$\Delta$ EoS. The value of the compactness at which the scalar charge appears is
almost the same for the cases of N, NH, and NH$\Delta$ matter. The value of $\omega$ increases by increasing the star compactness, with a maximum scalar charge for a special value of the compactness. Fig.~\ref{charge1} shows that $\omega$ then decreases by increasing $M/R$. With the value of $\beta=-4.5$ for the coupling constant, the scalar charge reaches zero in stars with high enough compactness. However, with the values of $\beta=-6$ and $\beta=-5$, the scalar charge is nonzero even in high compactness stars. For the case of $\beta=-4.5$, in the NSs with NH and NH$\Delta$ matter,
at a lower value of the compactness, the star scalar charge gets zero. The effects of the NH and
 NH$\Delta$ matter on the scalar charge are more significant for the stars with higher compactness. Fig.~\ref{charge1} also verifies that in the neutrino diffusion, neutrino-transparent, and cold-catalyzed stages,
the scalar charge in stars with NH$\Delta$ matter can be different from the one with NH matter. In the neutrino diffusion and neutrino-transparent regimes, the $\Delta$-resonances reduce the scalar charge. This is while in cold-catalyzed regime, the $\Delta$-resonances raise the scalar charge in stars with high compactness. By increasing the coupling constant, the minimum value of the compactness that results in the nonzero scalar charge increases and the range of the star compactness that the scalar charge
exists is smaller.

\begin{figure}[h]
	\subfigure{}{\includegraphics[scale=0.55]{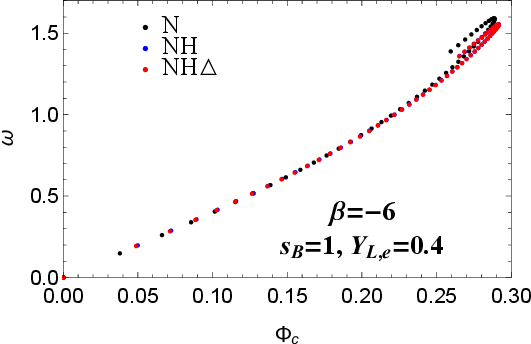}
		}		
\subfigure{}{\includegraphics[scale=0.55]{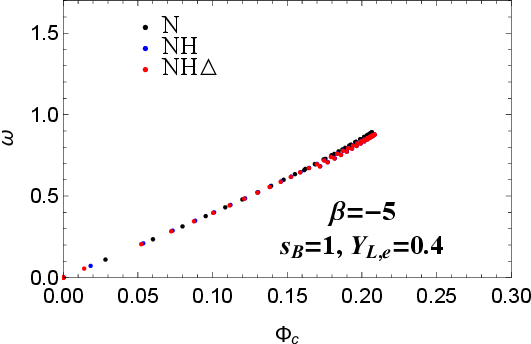}
		}	
\subfigure{}{\includegraphics[scale=0.55]{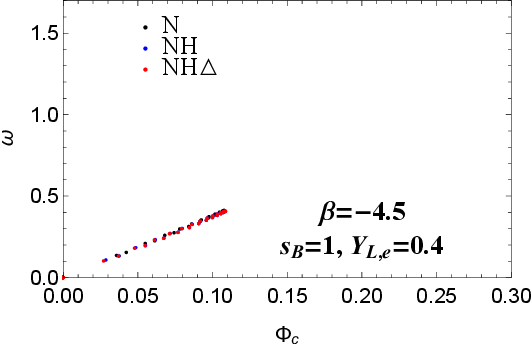}
		}	
\subfigure{}{\includegraphics[scale=0.55]{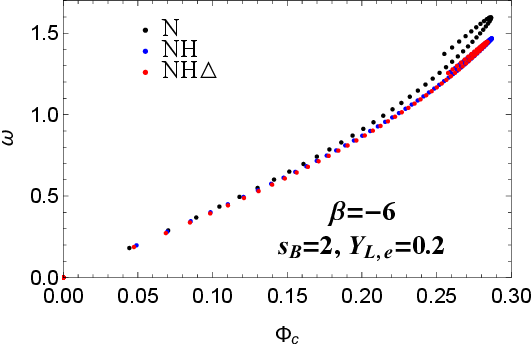}
		}		
\subfigure{}{\includegraphics[scale=0.55]{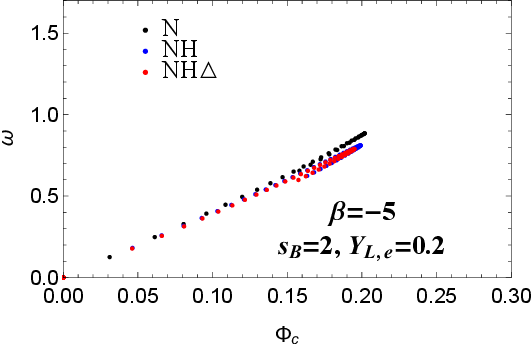}
		}	
\subfigure{}{\includegraphics[scale=0.55]{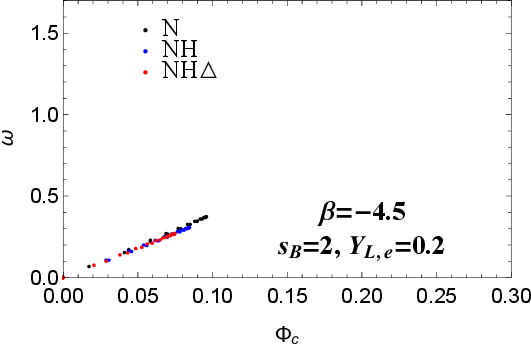}
		}	
	\subfigure{}{\includegraphics[scale=0.55]{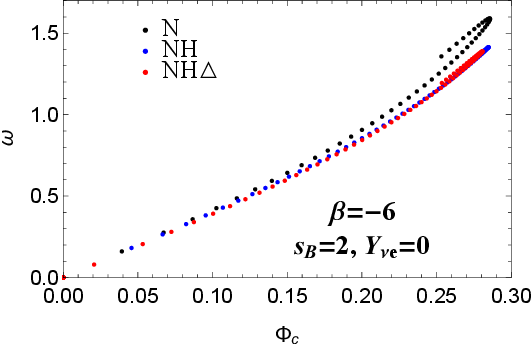}
		}		
\subfigure{}{\includegraphics[scale=0.55]{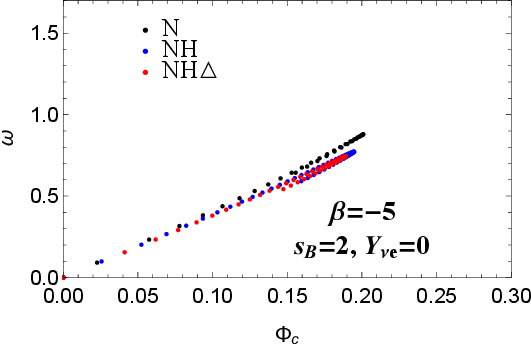}
		}	
\subfigure{}{\includegraphics[scale=0.55]{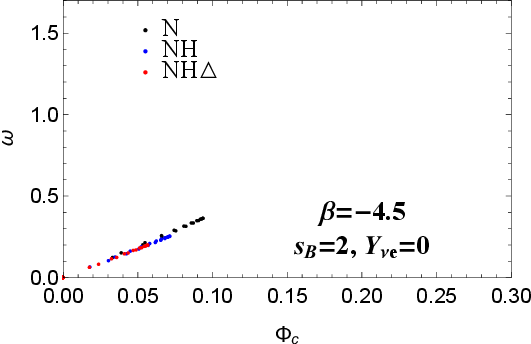}
		}
\subfigure{}{\includegraphics[scale=0.55]{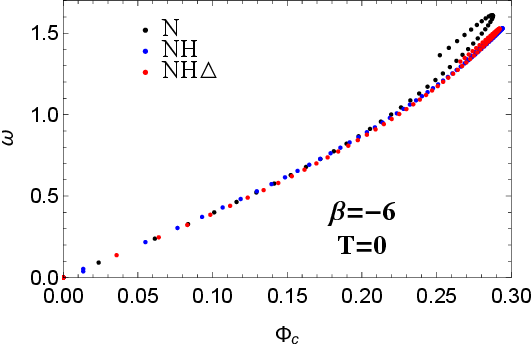}
		}		
\subfigure{}{\includegraphics[scale=0.55]{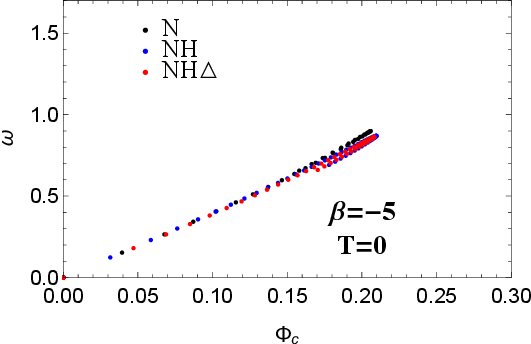}
		}	
\subfigure{}{\includegraphics[scale=0.55]{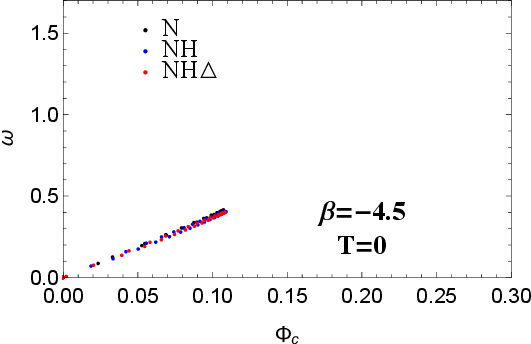}
		}	
	\caption{Scalar charge, $\omega$, versus the central scalar field, $\Phi_c$, in neutron stars with N, NH, and NH$\Delta$ matter and three values of the coupling constant $\beta$ considering the star's evolution, Neutrino-Trapped Regime ($s_B = 1, Y_L = 0.4$), Neutrino Diffusion Regime ($s_B = 2, Y_L = 0.2$), Neutrino-Transparent Regime ($s_B = 2, Y_{\nu_e} = 0$), and Cold-Catalyzed Regime (T=0).}
	\label{charge2}
\end{figure}

In Fig.~\ref{charge2}, we present the relation of the NS scalar charge with the central scalar field of the stars. In all cases, the scalar charge grows with $\Phi_c$. Moreover, the lower values of the coupling constant raise the rate of the scalar charge increase with $\Phi_c$. However, the presence of NH and NH$\Delta$ matter in NS reduces this rate when the central scalar field is high. The effects of NH and NH$\Delta$ matter on the NS scalar charge are more notable in the neutrino diffusion and neutrino-transparent regimes. With the lower values of $\beta$, both the central scalar field and scalar charge reach the larger values compared to higher values of $\beta$. Besides, NH and NH$\Delta$ matter affect the scalar charge more significantly in the cases with the lower values of $\beta$.

\section{SUMMARY AND CONCLUDING REMARKS}\label{s5}

In this paper, we explored the hot neutron stars with hyperons and $\Delta$-resonances during the star evolution in the scalar-tensor gravity.
We apply the equations of state for the stellar matter from the birth of a neutrino-rich proto-neutron star, neutrino-trapped star, neutrino diffusion star, neutrino-transparent star, and cold-catalyzed star. The equations of state were calculated in the relativistic mean-field approximation
for nucleons only (N), nucleons plus hyperons (NH), and nucleons plus hyperons plus $\Delta$-resonances (NH$\Delta$). Our results show that
the softening of the EoS due to the presence of hyperons and $\Delta$-resonances affects the range of the star scalarization. In the cases of NH and NH$\Delta$ stars in the neutrino diffusion, neutrino-transparent, and cold-catalyzed regimes, the scalarized stars can exist with higher central densities and
the range of the scalarization is more extended compared to the neutrino-trapped regime. Considering the neutrino diffusion, neutrino-transparent, and cold-catalyzed regimes, the second critical density of scalarization increases by the hyperons and decreases by the $\Delta$-resonances. The softening of the NH and NH$\Delta$ EoS also leads to the reduction of the star scalar charge. Our results underscore the critical role of exotic degrees of freedom (hyperons and $\Delta$-resonances) and thermal evolution in shaping the scalarization of neutron stars within scalar-tensor gravity. The softening of the EoS by these particles not only alters the density thresholds for scalarization but also modifies the star's mass radius relation and scalar charge, with pronounced effects in late-stage evolution (e.g., neutrino-transparent and cold-catalyzed regimes). These findings highlight the importance of incorporating finite-temperature effects and particle composition when testing modified gravity theories against multi-messenger observations, such as gravitational waves from mergers or NICER radius measurements. Future work could extend this framework to include quark deconfinement phases, rotation, magnetic fields, or dynamical scalarization in binary systems, offering a more complete picture of scalarized neutron star dynamics. Investigating how these ingredients influence observable quantities, such as gravitational waveforms, tidal deformabilities, and cooling profiles, and comparing them with data from next-generation observatories like the Einstein Telescope, Cosmic Explorer, and eXTP will be essential for constraining both the fundamental theory of gravity and the microphysics of dense matter.


\acknowledgements{Z.R. and F.R. wish to thank the Shiraz University Research Council. A.I. would like to thank the S\~ao Paulo State Research Foundation (FAPESP) for financial support through Grant No. 2023/09545-1.}

\end{document}